\documentclass[a4paper,english,12pt,amssymb,nofootinbib,superscriptaddress]{revtex4-2}
\usepackage{tcolorbox}
\usepackage{lmodern}
\usepackage{lmodern}

\usepackage[T1]{fontenc}
\usepackage[latin9]{inputenc}
\usepackage{babel}
\usepackage{mathrsfs}
\usepackage{amsmath}
\usepackage{amssymb}
\usepackage{esint}
\usepackage[unicode=true,pdfusetitle,
bookmarks=true,bookmarksnumbered=false,bookmarksopen=false,
breaklinks=false,pdfborder={0 0 1},backref=false,colorlinks=false]
{hyperref}

\makeatletter

\@ifundefined{textcolor}{}
{%
	\definecolor{BLACK}{gray}{0}
	\definecolor{WHITE}{gray}{1}
	\definecolor{RED}{rgb}{1,0,0}
	\definecolor{GREEN}{rgb}{0,1,0}
	\definecolor{BLUE}{rgb}{0,0,1}
	\definecolor{CYAN}{cmyk}{1,0,0,0}
	\definecolor{MAGENTA}{cmyk}{0,1,0,0}
	\definecolor{YELLOW}{cmyk}{0,0,1,0}
}

\usepackage{babel}
\usepackage{babel}
\usepackage{babel}
\usepackage{babel}
\usepackage{babel}
\usepackage{babel}
\usepackage{babel}
\usepackage{babel}
\usepackage{babel}
\usepackage{babel}
\usepackage{babel}
\usepackage{babel}
\usepackage{babel}
\usepackage{babel}
\usepackage{babel}
\usepackage{babel}
\usepackage{babel}
\usepackage{graphicx}
\def\b{\begin{equation}}
\def\e{\end{equation}}

\@ifundefined{textcolor}{}{%
	\definecolor{BLACK}{gray}{0}
	\definecolor{WHITE}{gray}{1}
	\definecolor{RED}{rgb}{1,0,0}
	\definecolor{GREEN}{rgb}{0,1,0}
	\definecolor{BLUE}{rgb}{0,0,1}
	\definecolor{CYAN}{cmyk}{1,0,0,0}
	\definecolor{MAGENTA}{cmyk}{0,1,0,0}
	\definecolor{YELLOW}{cmyk}{0,0,1,0}
}

\usepackage{latexsym}\usepackage{bm}

\makeatother

\begin{document}
	\title{A Tribute to S. Deser: Conserved Quantities in Generic Gravity Theories }

	\author{Bayram Tekin}
	
	\email{bayram.tekin@bilkent.edu.tr}
	\affiliation{Department of Physics,\\
		Middle East Technical University, 06800 Ankara, T\"urkiye}
	\altaffiliation{Present address: Department of Physics, Bilkent University, 06800 Bilkent, Ankara, T\"urkiye}

	\selectlanguage{english}%
\begin{abstract}
\noindent I describe parts of my joint work with S. Deser [March 19, 1931 - April 21, 2023] which started when I was working as a post-doc at Brandeis University in 2001.  Our work was mostly, but not exclusively, on conserved charges of higher curvature theories of gravity. I also describe some recent developments, such as expressing the conserved charges in terms of the linearized Riemann tensor; I expound upon the computations. I also reminisce about our interaction for over two decades that went beyond our scientific collaboration. The physics part of this review is intended for graduate students and researchers interested in extended theories of gravity, especially about the conserved quantities and perturbative techniques in these theories. 
\end{abstract}
\maketitle
\[
\]
\tableofcontents{} 
\[
\]

\section{A historical note about my interaction with S. Deser}

Tributes to well-established scientists like Deser are common; what is perhaps less common is for one to be written by a little-known physicist like myself, hence my hesitation. In the end, I decided that my vantage point in this interaction would do no harm and perhaps would even be of some benefit to some people. The age gap between Stanley and myself was over 40 years, so I can in no way claim that I knew him well. When we first shook hands, he was 70 and I was 28. 
Nevertheless, we interacted frequently: I have been to his house and his wonderful wife Elsbeth (Oskar Klein's daughter) cooked dinner for us, and we have kept good communication in the last two decades. In the first decade, we talked mostly about physics, and in the last one, our conversation centered around the evaporation of the trace amount of democracy we used to have in Turkey. Based on his own experience before and during WWII, he was urging me to leave. But here writing about that subject in more detail would offer me an undesired visit to a prosecutor's office. Instead, I will write a little bit about the two years I spent at Brandeis. Those readers who are more eager about physics are invited to skip this section.\footnote{This section also appears in \cite{KaneShifman}.}

\subsection{ There is something in a name. ``Write my name as S. Deser!''}

In September of 2001, when I finished writing our first joint work with Deser that summarized our talks on the board in my office, I wrote  ``Stanley Deser''  as an author and printed a copy for him to suggest corrections.  To be perfectly honest, the paper was not a very impressive one, but still, I was quite proud to have produced something publishable with Stanley whose works in the Hamiltonian formulation of gravity, supergravity, topologically massive gravity, and gauge theory had been written in stone.  I had turned down a potentially much more fruitful post-doc offer and came to Brandeis just to work with Stanley.  So joint work with him was exciting for me. 
After a glance at the cover page of the manuscript, he immediately requested me to shorten his name to ``S. Deser''  and noted that he had always signed all his works as such. I did not pay much attention to this incident and wrote his name as S. Deser in our joint works \cite{1,2,3,4,5,6,7,8,9,10,11,12}.
 At the time his insistence on writing S. seemed like some kind of representation of humility for me. My understanding was this: as an author, acknowledge that you are the author, but otherwise remove yourself from the room between the reader and the paper. For me on the other hand, as I come from a barren land as far as science is concerned, I explicitly write my name as there has been no other ``Bayram'' in gravity or high energy theory: B. would be too cryptic and could not have been guessed.  Back to S.: only recently did I come to understand his (almost) passion in shortening his name. My explanation could be poetic, but still, I think it could be true. In 2020 or 2021 Deser sent me an early draft of his book ``{\it Forks in the Road, A Life in Physics}'' \cite{forks}. I read the book immediately the night I received the draft and made some suggestions.  The book is a must-read. I learned from that book that he, as a Jewish boy, started his life first with the name ``Salomon''  in a Polish city (now in Ukraine) and left for Palestine with his name intact; but his name was changed to ``Lucien'' when the family moved to France when he was a refugee. His name was kept as Lucien when the family escaped to Spain and Portugal, but when he ended up in New York; his name was changed one more time to ``Stanley''. I think that old man that I worked with who insisted that his name be kept as S. Deser in the papers was always attached to the little Salomon, and that S. also referred to Salomon. The Nazis brought out the worst in humanity, forcing a boy to flee his name and home many times. This could happen again, it was not a one-time event in human history. We must be vigilant.

\subsection{September 11 2001: S. says ``We must evacuate!''}

On September 11, 2001, I was working in my office at the Brandeis University physics department. Stanley rushed into my effectively windowless office in a panic.
 He said ``I don't think  Palestinians can do something this big, but in any case, we have to evacuate the university''.  [Brandeis, a nonsectarian university founded by the American Jewish community, could have been seen as a symbolic target; in the initial hours of the September 11 attacks, it was not clear who was behind them or what else was a target.]
I had no idea what he was talking about, I had not heard the news. I did not know what happened. As he rushed out, I did not ask him.
 But I packed my bag and left and took the train to Boston from Waltham and went home.  That home was no home really: I had arrived in Boston from Oxford (UK) about 20 days earlier. I rented an empty house in Somerville. I was sleeping on the wooden floor. I had no phone, no computer, no TV, and no money. I had no idea what was going on in the USA, even though Boston was in some sense at the epicenter of the events. But clearly, something big was taking place. As the two planes that hit the World Trade Center had taken off from Boston's Logan Airport, something interesting happened: my landlord started to check on me. As a Middle Eastern-looking man, living in his house, with zero furniture, no bed, with just a large duffle bag and weird papers scattered over the floor, I sure was a suspect. He did not know that a poor theorist post-doc lived in less fortunate conditions than a well-funded terrorist. My salary was \$2000 per month while my rent was \$1400. 
Anyway, I decided to go to Harvard Square to understand what was going on. On the subway train, someone shouted that everybody should follow him and sing ``God Bless America''. That was interesting, I did not know the lyrics, I had lived in Oxford (UK) and had a slight inclination to sing ``God Save the Queen''.  
I learned about the terrible situation in Harvard Square and immediately realized the consequences: America will fall into the nationalism trap that I am all too familiar with from my own country which is de facto the landlord of that trap. 

Back to Stanley. When I applied for post-doc jobs as a graduate student in Minnesota, Stanley offered me a position to start in September 1999. But, I went to the UK. After two years, amazingly, Stanley was kind enough to offer me again.
 I arrived in Boston at about the end of August 2001.
 Stanley arranged a guestroom at Brandeis for me to stay just for a week. He was on vacation in Sweden. I started working. When he got back, after about a couple of minutes of introduction, he asked what I was working on. I explained on the board in the office my idea on Newtonian limits of massive gravity with a cosmological constant. Anyway, before the end of September, the paper was ready. And I told that story above. 

\subsection{Time Travel: S. says ``good luck!''  }

Sometime in 2002, I was at work in Stanley's office.  We were discussing, perhaps fine-tuning a paper that we were writing. His office phone rang. A journalist (perhaps, from the Boston Globe) was calling to ask about time travel in physics. 
What was the reason for such a call?    A  physics professor from the University of Connecticut had claimed recently that time travel was possible and he had been building a time machine. The journalist wanted to know what Deser thought of this. Deser with Roman Jackiw (sadly, Jackiw, a giant of physics, passed away 2 months after Stanley) had written a paper on time travel \cite{DJtime} and I guess the journalist found out about that paper. 
Anyway, Deser at one point on the phone mentioned me to the journalist.  He said, ``My post-doc Dr. Tekin is here, if you want you can talk to him about this''. Of course, the journalist did not want that. Anyway, in the end, Deser said ``Time travel requires organizing all the mass 
in the universe in a very particular configuration. So good luck!''. 
He may have had in mind G\"odel's solution of Einstein's equations \cite{Godel}, the so-called G\"odel universe: a rotating dust universe, in which closed timelike curves pass through every point, so that time travel to the past is allowed. G\"odel wrote the paper for the volume celebrating Einstein's 70th birthday. More likely, given his own work, he was alluding to the 1992 analyses of Gott's cosmic-string time machine \cite{DJtH,CFG,tHooft92}: in an open universe the total momentum of such a configuration turns out to be tachyonic, and in a closed universe the system collapses before a closed timelike curve can form. In either case, all the matter of the universe has to cooperate. 
If I had picked up the phone, I would have told that journalist the following: you want to go back in time in a relatively cheap way, just travel east, say to North East Anatolia, where I was born. You would be going back about 2-3 centuries.

\subsection{Deser's Wit and Languages }

Stanley was a witty and humorous person; he, most of the time, would not say something directly but instead, encase his message in a terse joke. One had to know him a little bit to understand what he was saying. I have hundreds of emails from him in my mailbox, and some of them made me emotional as I went through them while writing this. He tried to help me out when I was struggling with the authorities here. He offered his help several times and searched for possible solutions.   Some of his messages were just funny. For example here is one: just a message in the subject line and no text in the email  `` Thanks for the great ackn. BUT do NOT move in w.your collaborators!''. This was referring to a review I wrote with my student and two colleagues.  

One day Stanley brought me two preprints one in French and one in German. I do not remember exactly what we were discussing. He asked me to read them. I said that I did not read French and I could barely read German with a good dictionary. He seemed surprised because he spoke many languages and asked why  I did not know these languages.  I said that I came from a region where people barely spoke one language and learning a little bit of English was considered already a success.

Stanley was amazingly quick with his response to emails, one would have thought that he was always sitting in front of his computer. He was not so patient with some people that omitted some obvious references. But, since he would write his messages in the subject line with shortened words, very often the receivers would be baffled and ask questions as to what he meant. That usually turned into an interesting, rather slowly converging, exchange with Stanley answering in shortened words and hard-to-decipher sentences for those who did not know his style.

I must say this: when he did not like the introduction that I wrote for a paper, he wrote his version on some yellow paper with a stylographic pen. When he gave his version to me and I looked at it for a second and before I said anything he said ``Of course you cannot read it, but the secretary can.''  Stanley's handwriting was simply unreadable.  I remember him writing the address on an envelope that we were sending to the Class. Quantum Grav. journal and I was thinking to myself ``There is no way this mail will arrive there.''

As it could be common with his generation, corrections on an accepted paper never ended with Stanley: I remember that {\it Phys. Rev. Lett.} accepted our paper without any correction, but Stanley started a marathon of corrections that seemed to have no end. When I wrote a slightly angry email about this (as he was in Sweden at the time), he took his time to reply. After 3 days of silence, he replied with a very kind email and urged me to ask Roman Jackiw about his way of writing. Of course, I did not do that. 

Finally, let me note the following: our major joint work, conserved charges in generic gravity theories,  was summarized in \cite{3,4}. The work started as follows: Stanley brought the paper \cite{strom} to me and said that he thought the paper was not correct. This is a very interesting paper about purely quadratic gravity in four dimensions where it was shown that ``exact solutions representing isolated systems have precisely zero energy'' and so there is a confinement of energy in the purely quadratic theory.  Stanley could not accept such a confinement of energy in the sense that certain localized energy-momentum tensors could not affect the spacetime at larger distances. For him, there could be no confinement of energy. For me, it was possible.
 As per Stanley's request, I started working on how one might consistently define global charges in quadratic gravity theories for asymptotically flat and constant curvature spacetimes. Our first results were published as \cite{3} and a more detailed version was published in \cite{4} where we included a section discussing the zero energy issue which was the original impetus for that work.

What was our common interest that Stanley offered me a post-doc position? I never asked him this question, but I have a guess. I did my PhD in Minnesota between the years 1994 and 1999 working with my advisor Yutaka Hosotani in various perturbative and non-perturbative problems in quantum field theory. I learned a great deal from Yutaka, especially in tackling hard and long computations. Minnesota had remarkable experts on quantum field theory and particle physics. But nobody was working on gravity. While working on the confinement problem in the 2+1 dimensional Georgi-Glashow model with a Chern-Simons term, I  got hooked on a paper Witten wrote \cite{Witten2} on 2+1 dimensional quantum gravity and its equivalence to Chern-Simons theory. So I started working on the subject; and submitted two manuscripts to the physics ArXiv on this before I applied for post-doc jobs.  I think Stanley offered me a post-doc job twice due to these works, and my main research shifted towards gravity. I am grateful to him for the subtle things I learned in our endeavor. We had several unfinished discussions with him on the degenerate vacua of gravity. He was hard to persuade and so I left those works unfinished. I will remember S. Deser for the rest of my life as a teacher and a colleague. 

\section{Introduction}

\subsection{ADM energy, linear momentum and angular momentum for asymptotically flat spacetimes}

For any {\it geometric} theory of gravity based on the Riemann tensor with a Lagrangian density $\mathcal{L}\equiv\sqrt{-g}\,f\left(R_{\rho\sigma}^{\mu\nu}\right)$, where $f$ is at least a twice differentiable function of the Riemann tensor, the total conserved energy of a 3+1 dimensional spacetime is given by the  Arnowitt-Deser-Misner
(ADM) \cite{adm} formula for \emph{asymptotically flat} spaces:\footnote{ Most of the formulas will be valid in generic $n$ spacetime dimensions, but for concreteness, we shall sometimes work in the $n=3+1$ dimensions. Also, if the spacetime has various asymptotically flat ends, then the ADM energy is valid for each end.}  
\begin{tcolorbox}
\begin{eqnarray}
E_{ADM}=\frac{1}{2\kappa}\int_{S_\infty^{2}} dS\, n^i  \Big (\partial_{j}h_i^j-\partial_{i}h_j^j\Big),\label{eq:ADM_e}
\end{eqnarray}
\end{tcolorbox}
\noindent where the ``deviation'' from the flat metric is defined as $h_{\mu\nu}\equiv g_{\mu\nu}-\eta_{\mu\nu}$ and requires proper decay conditions to satisfy the notion of asymptotic flatness in these coordinates. Naively the above integral should make sense: it should be zero for a flat space and non-zero for a curved one. (As we note in Sec. II B, this naive expectation can fail in badly chosen coordinates.) The integral is to be evaluated on a sphere at spatial infinity and $ n^i= \frac{x^i}{r}$; and in the units of $G_N=1=c$, $\kappa = 8 \pi$.
Even though, the formula is written in terms of Cartesian coordinates, it is a {\it geometric invariant } of the spatial part of the spacetime manifold, modulo proper decay assumptions.   This geometric invariance property attracted a great deal of research on the subject from the vantage point of differential geometry: a geometrically invariant quantity (mass-energy) can be defined for any asymptotically (locally) Euclidean three-manifold under certain conditions without referring to the full spacetime. Thus $E_{ADM}$ can be studied as a purely geometric object and one ends up with the famous positive mass theorem \cite{Desersugra,Schoen,Witten} where it was shown that $E_{ADM} \ge 0$ which is not apparent at all from  (\ref{eq:ADM_e}).

We can also define a total ADM linear momentum of an asymptotically flat spacetime as
\begin{equation}
P_{i}=\frac{1}{\kappa}\int_{S^2_\infty} dS\, n^j\, \Big(K_{ij}-K\gamma_{ij}\Big),\hskip 1 cm K:=\gamma^{ij}K_{ij},
\label{mom}
\end{equation}
where $ K_{ij}$ is the extrinsic curvature of the 3-dimensional spatial hypersurface in spacetime: it essentially describes how the spatial manifold is embedded in spacetime. In the usual ADM decomposition of the 4-dimensional manifold in terms of the lapse function $N=N(t,x^{i})$ and the shift vector $N^{i}=N^{i}(t,x^{i})$
and the spatial Riemannian metric $\gamma_{ij}= \gamma_{i j}(t, x^j)$
\begin{equation}
ds^{2}=(N_{i}N^{i}-N^{2})dt^{2}+2N_{i}dtdx^{i}+\gamma_{ij}dx^{i}dx^{j},
\hskip 1 cm i, j \in (1,2,3),
\label{ADMdecompositionofmetric}
\end{equation}
the extrinsic curvature reads as  
\begin{equation}
K_{ij}=\frac{1}{2N}\Big(\dot{\gamma}_{ij}-D_{i}N_{j}-D_{j}N_{i}\Big), \hskip 1 cm \dot{\gamma}_{ij} = \frac{\partial}{\partial t}\gamma_{ij},
\end{equation}
where   $D_{i}$ is covariant derivative compatible with the spatial metric  $\gamma_{ij}$.\footnote{ From a more geometric point of view, the extrinsic curvature is defined as follows in a coordinate invariant manner: let $n$ be the unit normal to the spacelike hypersurface and $X,Y$ be two tangent vectors at a point $p$ in the hypersurface, and $\nabla$ be the spacetime-metric compatible connection, then $K(X,Y):=g(\nabla_{X}n,Y)$.} 

Having the extrinsic curvature at our disposal, we can also define the total angular momentum of an asymptotically flat 1+3 dimensional spacetime as\footnote{The trace part of $K^{kl}-K\gamma^{kl}$ drops out of (\ref{dad}) since $n_l \propto x_l$ on the sphere at infinity.}
\begin{equation}
J_{i}=\frac{1}{\kappa}\varepsilon_{i j k}\int_{S^2_\infty} dS\, n_l \,\Big ( x^j K^{k l} -  x^k K^{j l}\Big).
\label{dad}
\end{equation}
As we shall {\it derive} conserved charges in generic coordinates for both asymptotically flat and asymptotically constant curvature and even for asymptotically Einstein spacetimes, we have only given the expressions above without a derivation. Note that the angular momentum was not mentioned in the original ADM 
work, see the discussion in the wonderful book \cite{Eric}.

\subsection{Conserved charges for asymptotically constant curvature spacetimes}

The discussion above was for asymptotically flat spacetimes; the construction changes in a significant way for asymptotically (anti)-de Sitter {[}(A)dS{]} spacetimes: the conserved charges are no longer simply geometric invariants of the manifold, but they are \emph{theory-dependent} quantities.  For asymptotically Einstein spaces, the story is much more complicated.  Generically,  all the parameters of a theory that appear in the action contribute to the conserved charge expressions. For example,  let us note that while the asymptotically flat Kerr black
hole solution has the same mass and the same angular momentum in all geometric theories of gravity of which it is a solution (in four dimensions), its asymptotically
(A)dS version, the Kerr-(A)dS black hole has different, numerically \emph{scaled}
masses and angular momenta for each theory to which it is a solution. So if one considers asymptotically constant curvature vacua, 
all higher order tensors in the action will contribute to the conserved charges; and if this fact is neglected the first law of black hole thermodynamics cannot be satisfied.

Since at both low and high energies, General Relativity is expected to be modified for different reasons, one should build a procedure to construct conserved charges in a given higher derivative theory.
 What is perhaps also quite important is to find a formula that works in all coordinates not just a specific one.  Of course, some coordinates can be quite terrible, as discussed in detail in \cite{ST,book,review} in such terrible coordinates even empty flat spacetime may turn out to have any mass and angular momentum we assign. 
There are various ways to approach the construction of conserved charges in generic gravity theories, and trying to write a historically accurate review would be quite non-trivial and it is beyond the scope of this tribute.  Instead, we will take a more linear and pedagogical point of view here.

The first generalization of the ADM energy was given by Abbott and Deser (AD)
\cite{Abbott} in cosmological Einstein's gravity for asymptotically
(A)dS spacetimes.  Their major goal in that work was to understand the stability of de Sitter spacetime. That requires a definition of energy, which is subtle in de Sitter spacetime since it lacks a global time-like Killing vector field and a spatial infinity; the energy can be defined within the cosmological horizon.   But let us follow the AD construction and assume the existence of a background Killing vector $\bar{\xi}^\mu$, then
one has the following total conserved Killing charge,  known as the Abbott-Deser conserved charge
\begin{tcolorbox}
\begin{eqnarray}
Q(\bar{\xi})=\frac{1}{\kappa}\int_\Sigma dS_{i}\sqrt{-\bar{g}}\Big( \bar{\xi}_{\nu}\bar{\nabla}_{\beta}K^{0i\nu\beta}-K^{0j\nu i}\bar{\nabla}_{j}\bar{\xi}_{\nu}\Big) .
\end{eqnarray}
 The ``superpotential'' $K^{\mu\alpha\nu\beta}$ is defined as
\begin{eqnarray}
K^{\mu\alpha\nu\beta}:=\frac{1}{2}\Big (\bar{g}^{\mu\beta}\tilde h^{\nu\alpha}+\bar{g}^{\nu\alpha}\tilde h^{\mu\beta}-\bar{g}^{\mu\nu}\tilde h^{\alpha\beta}-\bar{g}^{\alpha\beta}\tilde h^{\mu\nu}\Big ),\hskip0.7 cm \tilde h^{\mu\nu}:=h^{\mu\nu}-\frac{1}{2}\bar{g}^{\mu\nu}h.
\label{sup}
\end{eqnarray}
\end{tcolorbox}
\noindent The tensor $K^{\mu\alpha\nu\beta}$ has the algebraic symmetries of the Riemann tensor.\footnote{In a more compact notation, we can write the $K$ tensor as a Kulkarni-Nomizu product $ K= - \frac{1}{2}\bar g \mathbin{\bigcirc\mspace{-15mu}\wedge\mspace{3mu}} \tilde h$, with the convention $(A\mathbin{\bigcirc\mspace{-15mu}\wedge\mspace{3mu}}B)_{\mu\alpha\nu\beta}=A_{\mu\nu}B_{\alpha\beta}+A_{\alpha\beta}B_{\mu\nu}-A_{\mu\beta}B_{\alpha\nu}-A_{\alpha\nu}B_{\mu\beta}$.} The superscript $0$ in the integrand already tells us that one has assumed the existence of a global time-like direction (at least outside a compact domain where there could be a black hole region). Otherwise, a conserved charge would not make sense as one would not have $\frac{d Q}{dt}=0$.  But this is not surprising: one does not have conserved charges for a generic spacetime without asymptotic symmetries.  The AD expression can be used for asymptotically constant curvature spaces but it is also valid for {\it asymptotically Einstein} spaces as we shall show below.   In the notation of \cite{3,4}, the AD expression reads (in any spacetime dimension) as 
\begin{eqnarray}
Q_{\text{Einstein}}^{\mu}(\bar{\xi})=\frac{1}{2\kappa}\int_{\Sigma} & dS_{i} & \Big (\bar{\xi}_{\nu}\bar{\nabla}^{\mu}h^{i\nu}-\bar{\xi}_{\nu}\bar{\nabla}^{i}h^{\mu\nu}+\bar{\xi}^{\mu}\bar{\nabla}^{i}h-\bar{\xi}^{i}\bar{\nabla}^{\mu}h\nonumber \\
 &  & +h^{\mu\nu}\bar{\nabla}^{i}\bar{\xi}_{\nu}-h^{i\nu}\bar{\nabla}^{\mu}\bar{\xi}_{\nu}+\bar{\xi}^{i}\bar{\nabla}_{\nu}h^{\mu\nu}-\bar{\xi}^{\mu}\bar{\nabla}_{\nu}h^{i\nu}+h\bar{\nabla}^{\mu}\bar{\xi}^{i}\Big).\label{ad1}
\end{eqnarray}
For $\mu=0$, $Q^{0}(\bar{\xi})$ gives the corresponding
energy or angular momentum once the background Killing vector $\bar{\xi}^{\mu}$
is specified. {[}Note that $Q^{i}\left(\bar{\xi}\right)$ is the total current.] 

As stated above, what is quite satisfying about (\ref{ad1})
is that it not only works for asymptotically (A)dS spacetimes but also for asymptotically flat ones as well as asymptotically Einstein spacetimes in an \emph{arbitrary} coordinate system with the condition that the coordinates  are well-behaved at infinity,
see the Appendix of \cite{ST}.

\subsection{Higher curvature theories of gravity }

We shall study theories beyond General Relativity\footnote{This section is adapted from Chapter 9 of the book \cite{book}. I thank my colleagues for allowing me to re-use the material.},  therefore it first pays to motivate why such modifications are needed at all.   As far as our understanding of the universe is concerned, we are living in the era of ``effective field theories'',  this is because we have learned that all theories we have built so far; or we shall build are valid up to some energy scales at which new physics, new degrees of freedom and perhaps new symmetries,  enter into the picture and our most current theory needs to be modified. So from this vantage point, there is a lot of motivation to study the generalizations of  Einstein's gravity with an action symbolically written in the  form
\begin{equation}
S=\int d^{n}x \sqrt{-g}\,\Bigg ( \frac{1}{\kappa}\left(R-2\Lambda_{0}\right)+\sum_{p=2}^{\infty}a_{p}\Big(\text{Riem, Ric, R, }\nabla\text{Riem, }\dots\Big)^{p}\Bigg).\label{generic_higher}\end{equation}
\noindent A restricted version of these theories, the quadratic conformally invariant theory based on the square of the Weyl tensor dates back to the early days of General Relativity.   Hermann Weyl argued that the correct framework of unifying gravity and electromagnetism was to allow conformal scalings of the metric as a symmetry of the underlying theory.  In such a theory, the gravity sector is represented not by the Einstein-Hilbert term but by the square of the Weyl tensor.  We know that this idea does not work in the original incarnation as it allows history dependent spectrum for atoms, which was not observed, but its basic premise-the gauge principle- is omnipresent in all modern physics.

Higher order terms in the action  (\ref{generic_higher}) can arise as a result of integrating out massive degrees of freedom in a microscopic theory, such as in string theory, or one might simply take general terms consistent with diffeomorphism invariance and build a phenomenology of the resulting theory. There could be other fields, such as gauge fields, and scalar fields that non-minimally couple to gravity, but for now let us not consider them. The non-minimally coupled scalar field brings about non-trivial issues; we shall only point to the literature on it in the Conclusions. This type of higher derivative theory, with propagators that decay faster than $1/p^2$ at large momenta, has much better behavior in the ultraviolet regime and reduces to General Relativity (GR) at large distances (in the infrared) bringing in only some weak constraints on the couplings $a_p$ from the solar system experiments.  But a better behavior in the ultraviolet regime does not mean that the theory is kosher: generically as a fundamental theory a higher derivative theory does not make sense as there are ghosts around and the unitarity is ruined \cite{Stelle1,Stelle2,Woodard}. We shall quote the particle spectrum of quadratic and $f(R^{\mu\nu}_{\rho\sigma})$ theories in Sec. V.\footnote{Incidentally, Stanley told me the following anecdote.  At the dinner table of a physics conference, Hawking argued in favor of ghosts and asked ``Who is afraid of ghosts?'', and Stanley answered `` I am''.  Hawking and Hertog wrote a paper titled `` Living with ghosts'' \cite{HH}.}

Two of the problems of GR arise at large distances: these are the problems of current or recent accelerated expansion of the universe (that is: what drives the current accelerated expansion of the universe?) and the rotation curves of stars in spiral galaxies (why stars far away from the bulge of the galaxy rotate around the center with constant speeds?). These large-scale problems require the introduction of dark energy and dark matter to GR in amounts that are far more than the visible matter and photons.  The following might turn out to be the solution to these problems: namely, the theory is not modified but is augmented with dark energy and dark matter yet to be detected by other means than gravity. Of course, it would still be a great mystery why most of the energy/matter budget of the universe is stuff that does not seem to exist in the standard model. There also arises the problem, or rather a curiosity that needs an explanation is the ``cosmic coincidence'' as to why dark matter and dark energy are of the same order of magnitude. 

There is another possibility of solving the above-mentioned large-scale problems within pure gravity without introducing dark matter and dark energy: one such route is a recently resurrected, trend of giving a tiny mass to the graviton that not only weakens the gravity at large distances but almost acts like dark energy.  [Admittedly, this route is a very cumbersome one in four dimensions and above since there is no natural geometric structure on a manifold that can be identified with graviton's mass in a non-linear setting.  We also do not know how symmetry breaking a la Higgs can generate graviton's mass.] Logically the third possibility could be a combination of both (which I favor): there may be some amount of dark matter and dark energy (whose nature and amount we still have to find) and also, the graviton has a tiny mass.  Here, we shall not be interested in the phenomenological aspects of large-scale gravity, we shall only discuss the massive modes in higher derivative theories. Massive modes in higher curvature gravity theories exist, because, generically, once higher curvature terms are added to the Einstein-Hilbert action, besides the massless spin-2 particle of General Relativity, new degrees of freedom, which are generically massive, arise.  For example, massless GR in $n$ dimensions has $n(n-3)/2$ degrees of freedom, while GR augmented with the Fierz-Pauli mass term has $(n+1)(n-2)/2$  degrees of freedom.  There is also a third possibility in the presence of a cosmological constant $\Lambda$,  if the mass term and the cosmological constant are tuned as $m^2= 2 \Lambda/(n-1)$, then there arises a new symmetry and one has a partially massless theory with one less degree of freedom than the massive one \cite{Hig,Waldron}.\footnote{Stanley and, at the time, his post-doc, Andrew Waldron did nice work on partially massless gravity theories and other field theories.}  Among higher curvature theories, quadratic gravity plays a particularly important role and hence we shall spend time on figuring out its particle content exactly and construct its conserved quantities. Once quadratic gravity is understood, a large class of generic theories constitutes a rather straightforward generalization of it.

In the other extreme, at high energies or extremely small (microscopic) scales,  there is no pressing experimental result that forces us to strictly modify GR. Of course, this should not deter us from such attempts because there are some compelling theoretical problems about GR in the UV scales:  pure GR (without matter) is non-renormalizable at the two-loop level in perturbation theory \cite{GS}, while with matter, the state of affairs is even worse: it is non-renormalizable even at the one-loop level \cite{tHV,DvN1,DvN2}.  So as a quantum theory, GR is only reliable at the tree level, namely at the level of a graviton exchange between conserved sources. The statement that GR is non-renormalizable is sometimes misunderstood:  it does not mean that the theory cannot be made finite it can be made finite at each order in perturbation theory including the loops, but the theory loses its predictive power at high energies, since one requires infinitely many couplings to be made finite or renormalized.  Hence it requires an infinite number of different measurements or data so
  GR is at best a valid effective theory at low energies.  

There is currently no viable quantum version of a higher derivative theory based on the metric and additional fields (supersymmetric or not), even though there is always hope about maximal supergravity theory being perhaps divergence-free at the several loop level (the possible appearance of supergravity counterterms at three loops was first pointed out in \cite{DKS}), which in any case is not good enough, the theory should be divergence-free at all loops.  Needless to say, when one deviates from the idea that the metric is the microscopic field and accepts that the spacetime, the metric {\it etc.}  are  {\it emergent } low energy quantities, one can build a renormalizable gravity theory.  String theory is a unique example with a perturbatively valid theory of quantum gravity where the metric is not the fundamental field but appears a posteriori. Namely, gravity, as we know it becomes a low-energy phenomenon appearing to our vulgar eyes. What we are interested in here is possible deformations of Einstein's gravity in the form (\ref{generic_higher}) and sometimes in the form  $f(R^{\mu\nu}_{\sigma\rho})$ that are better effective theories at high energies compared to GR.  So, we assume that at high energies gravity is still defined within the context of Riemannian geometry, with perhaps additional scalar fields.
 As mentioned, among these theories, quadratic gravity plays a special role as it is a renormalizable theory in four dimensions \cite{Stelle1}, but unfortunately, there is a massive spin-2 ghost in the spectrum about both the flat and the constant curvature vacua, so that neither vacuum is stable once the ghost is excited, which is of course unacceptable. (For attempts to make sense of the theory despite the ghost, see, e.g., \cite{DM,Salvio}.) Bartering unitarity with renormalizability is not a good deal for a physical theory, because the theory is a predictive one which is good but it predicts nonsense in the form of negative probabilities or higher than unity probabilities, which are bad. Thus, in this sense, generic quadratic gravity is not acceptable. One exception could be the 3-dimensional toy model with a special tuning of the quadratic terms or its infinite order extension which we shall briefly discuss.  Less generic gravity theories such as the one with the Lagrangian density  ${\cal{ L}} = R + \alpha R^2$  can be unitary but not renormalizable,  but
 of course they are not completely useless. For example, this particular theory leads to a successful inflationary phase in the early universe that is consistent with the cosmological observations at this stage. The theory is called the Starobinsky model \cite{Staro} and is equivalent to Einstein's gravity coupled to a self-interacting scalar field with a specific interaction potential.

\subsection{Conserved charges in quadratic theories of gravity}

A further generalization of the ADM expression was carried out for asymptotically (A)dS backgrounds in \cite{3,4}\footnote{The motivation for this work was given in Sec. I.}
for quadratic gravity theory with the action

\noindent 
\begin{equation}
I=\int d^{n}x\,\sqrt{-g}\left[\frac{1}{\kappa}\left(R-2\Lambda_{0}\right)+\alpha R^{2}+\beta R^{\mu\nu}R_{\mu\nu}+\gamma\left(R^{\mu\nu\rho\sigma}R_{\mu\nu\rho\sigma}-4R^{\mu\nu}R_{\mu\nu}+R^{2}\right)\right].\label{eq:Quadratic_action}
\end{equation}
The $\gamma$ term was organized into the Gauss-Bonnet combination\footnote{Let us make a side remark. There has been a great deal of activity on the 4-dimensional Einstein-Gauss-Bonnet theory in recent years. Let us quote M. Perry  ``The intellectual history of general relativity is littered with corpses of people who tried to manipulate the Euler character into the action of general relativity'' from  M. Perry 2007 Lecture notes. Despite our efforts \cite{gb1,gb2}, we were not able to put a complete halt on this non-existent theory in 4 dimensions. } which is a topological invariant for 4-dimensional compact manifolds hence does not contribute to the field equations in that dimensions; and vanishes identically below 4 dimensions. The procedure of finding the conserved Killing charges in this theory and a generic theory of gravity will be given in detail below. Here we just quote the final result for asymptotically AdS spacetimes: 
\begin{align}
Q_{\text{quadratic}}^{\mu}(\bar{\xi})= & \left(\frac{1}{\kappa}+\frac{4\Lambda n}{n-2}\alpha+\frac{4\Lambda}{n-2}\beta+\frac{4\Lambda\left(n-3\right)\left(n-4\right)}{\left(n-1\right)\left(n-2\right)}\gamma\right)Q_{\text{Einstein}}^{\mu}(\bar{\xi}),\label{eq:Quad_charge}
\end{align}
where $Q_{\text{Einstein}}^{\mu}(\bar{\xi})$ is given in (\ref{ad1})
(but without its overall factor $1/(2\kappa)$, i.e., with $2\kappa=1$).  There are some subtle issues to discuss: for example, here  the ``effective'' cosmological
constant $\Lambda$ satisfies the quadratic equation
\begin{equation}
\frac{\Lambda-\Lambda_{0}}{2\kappa}+\left[\left(n\alpha+\beta\right)\frac{\left(n-4\right)}{\left(n-2\right)^{2}}+\gamma\frac{\left(n-3\right)\left(n-4\right)}{\left(n-1\right)\left(n-2\right)}\right]\Lambda^{2}=0.\label{quadratic}
\end{equation}
So in four dimensions (A)dS, i.e. the vacuum, with zero charges, is unique with $\Lambda = \Lambda_0$, in quadratic gravity, but above four dimensions,  there are generically 2 vacua for the quadratic theory and $p$ vacua for a theory with curvature terms up to order $p$. There is no way to choose one effective cosmological constant over another one at this stage: these vacua should be considered as different universes with different asymptotic structures. Their symmetry group are the same but the numerical value of their constant curvature differ.  Of course, as $p$ gets large in a microscopic theory of gravity such as in string theory,  the number of these vacua increases which is an important problem to be addressed, but this problem is beyond the scope of this current review. On the other hand, we can mention some Born-Infeld-type higher curvature theories with infinitely many powers of curvature that have a unique vacuum \cite{b1,b2,b3,b4}.

\section{Detailed Calculations of the Deser-Tekin paper \cite{4}}

This section will be a detailed exposition of \cite{3,4}.  We will use mostly the same notation for ease of comparison. Our conventions are as follows: we use the mostly plus signature and the Riemann and the Ricci tensors are defined as $\left[\nabla_{\mu},\nabla_{\nu}\right]V_{\lambda}=R_{\mu\nu\lambda}\,^{\sigma}V_{\sigma}$,\,\,
$R_{\mu\nu}\equiv R_{\mu\lambda\nu}\,^{\lambda}$.

Let $R$ denote the Riemann tensor or its non-trivial contractions; $\nabla$ denote a metric-compatible, torsion-free connection; and start with a generic gravity theory, coupled to
a covariantly conserved, bounded, matter source $\tau_{\mu\nu}$ 
\begin{eqnarray}
\Phi_{\mu\nu}(g,R,\nabla R,R^{2},...)=\kappa\tau_{\mu\nu},\label{generic-1}
\end{eqnarray}
 where $\Phi_{\mu\nu}$ is the ``Einstein tensor\textquotedbl{} of local, invariant, but otherwise arbitrary, gravity action and $\kappa$
is the coupling constant. So we have the ``Bianchi Identity'', $\nabla^\mu \Phi_{\mu\nu}=0$, for all smooth metrics. 

In some suitable coordinates, let us split the metric   as
\begin{eqnarray}
g_{\mu\nu}=\bar{g}_{\mu\nu}+\kappa h_{\mu\nu},
\label{decompo}
\end{eqnarray}
where the ``background'' $\bar{g}_{\mu\nu}$ is assumed to solve (\ref{generic-1})
for $\tau_{\mu\nu}=0$; and $h_{\mu\nu}$ is the deviation
that vanishes sufficiently rapidly at (spatial) infinity,  but it need not be small in the interior region.  Any barred tensor will refer to the background geometry.
We introduced the coupling constant $\kappa$ in the splitting of the metric to count the order of the perturbation expansion.  
The effects of changing the coordinates on the decomposition (\ref{decompo}) leads to a change in the splitting, but for infinitesimal coordinate changes 
\begin{equation}
x^\mu \rightarrow x^\mu - \kappa \zeta^\mu(x),
\end{equation}
the deviation changes as 
\begin{equation}
h'_{\mu \nu}(x) = h_{\mu \nu}(x) + \bar\nabla_\mu \zeta_\nu +\bar\nabla_\nu \zeta_\mu,
\end{equation}
 which is the infinitesimal gauge transformation (diffeomorphism) acting on the non-invariant gauge field, usually expressed as the Lie derivative $\mathcal{L}_\zeta g_{\mu \nu}$. 
Tensors depending on the metric split  according to this decomposition and in particular, the field equation (\ref{generic-1})  becomes 
\begin{eqnarray}
\bar\Phi_{\mu\nu}(\bar{g},\bar{R},\bar{\nabla}\bar{R},{\bar{R}}^{2}...)+\kappa {\cal {O}}(\bar{g})_{\mu\nu\alpha\beta}h^{\alpha\beta}=\kappa T_{\mu\nu} (\tau, \kappa h^2, \kappa^2 h^3,...),\label{ope-1}
\end{eqnarray}
where the right-hand side includes all the terms except the zeroth and first-order term in the metric deviation plus the localized matter source which is assumed to be of the same order of magnitude as the deviation; while the left-hand side has the background and the linear order terms.  By assumption, the first term vanishes as the background metric satisfies the vacuum field equations.  The background operator ${\cal {O}}(\bar{g})$ is important: it is a formally self-adjoint operator in the sense that had we expanded the action of the theory, instead of the field equations, that term would have come from the variation of the second order expansion
\begin{equation}
-\frac{\kappa}{2 }\int _{\mathcal{M}}d^{n}x\sqrt{-\bar{g}}h^{\alpha\beta}{\cal O}\left(\bar{g}\right)_{\alpha\beta\mu\nu}h^{\mu\nu}.
\end{equation}
To understand this better, let us see the explicit form of this operator in cosmological Einstein's gravity when the background metric is maximally symmetric.  So we have the following classical vacuum (background) which will be assigned zero conserved charges:
\begin{equation}
\bar R_{\mu \alpha \nu \beta} = \frac{2 \Lambda}{(n-2)(n-1)} \left ( \bar g_{\mu \nu} \bar g_{\alpha \beta} -\bar g_{\mu \beta} \bar g_{\alpha \nu} \right ), \hskip 0.5 cm \bar R_{\mu \nu} = \frac{2 \Lambda} {n-2}\bar g_{\mu \nu},  \hskip 0.5 cm \bar R = \frac{2 n \Lambda}{n-2},
\end{equation}
which solves the vacuum Einstein equations 
\begin{equation}
\bar G_{\mu \nu} = \bar R_{\mu \nu} - \frac{1}{2} \bar g_{\mu \nu} \bar R + \Lambda \bar g_{\mu \nu} =0. 
\end{equation}
 For Einstein's gravity, one has the linearized Einstein tensor,  ${\cal G}_{\mu\nu}^{(1)}:= \big ( G_{\mu \nu} \big)^{(1)}$ which is  found to be 
\begin{equation}
{\cal G}_{\mu\nu}^{(1)}= R_{\mu\nu}^{(1)}-\frac{1}{2}\bar{g}_{\mu\nu}R^{(1)}-\frac{2}{n-2}\Lambda h_{\mu\nu},
\end{equation}
where, more explicitly one has 
\begin{align}
{\cal G}_{\mu\nu}^{(1)}= & \frac{1}{2}\left(-\bar{\Box}h_{\mu\nu}-\bar{\nabla}_{\mu}\bar{\nabla}_{\nu}h+\bar{\nabla}^{\sigma}\bar{\nabla}_{\nu}h_{\sigma\mu}+\bar{\nabla}^{\sigma}\bar{\nabla}_{\mu}h_{\sigma\nu}\right)\nonumber\\
 & -\frac{1}{2}\bar{g}_{\mu\nu}\left(-\bar{\Box}h+\bar{\nabla}_{\rho}\bar{\nabla}_{\sigma}h^{\rho\sigma}-\frac{2\Lambda}{n-2}h\right)-\frac{2}{n-2}\Lambda h_{\mu\nu}.
\end{align}
Therefore,  using  ${\cal G}_{\mu\nu}^{(1)}=:{\cal O}\left(\bar{g}\right)_{\mu\nu\alpha\beta}  h^{\alpha \beta}$, one can define the $(0,4)$ rank background operator as 
\begin{align}
{\cal O}\left(\bar{g}\right)_{\mu\nu\alpha\beta} := & \frac{1}{2}\left(-\bar{g}_{\mu\alpha}\bar{g}_{\nu\beta}\bar{\Box}-\bar{g}_{\alpha\beta}\bar{\nabla}_{\mu}\bar{\nabla}_{\nu}+\bar{g}_{\mu\alpha}\bar{\nabla}_{\beta}\bar{\nabla}_{\nu}+\bar{g}_{\nu\beta}\bar{\nabla}_{\alpha}\bar{\nabla}_{\mu}\right)\nonumber \\
 & -\frac{1}{2}\bar{g}_{\mu\nu}\left(-\bar{g}_{\alpha\beta}\bar{\Box}+\bar{\nabla}_{\alpha}\bar{\nabla}_{\beta}-\frac{2\Lambda}{n-2}\bar{g}_{\alpha\beta}\right)-\frac{2}{n-2}\Lambda\bar{g}_{\mu\alpha}\bar{g}_{\nu\beta},
\end{align}
which, after rearrangement is
\begin{align}
{\cal O}\left(\bar{g}\right)_{\mu\nu\alpha\beta}= & \frac{1}{2}\bar{g}_{\mu\nu}\bar{g}_{\alpha\beta}\left(\bar{\Box}+\frac{2\Lambda}{n-2}\right)-\frac{1}{2}\bar{g}_{\mu\alpha}\bar{g}_{\nu\beta}\left(\bar{\Box}+\frac{4\Lambda}{n-2}\right)-\frac{1}{2}\left(\bar{g}_{\mu\nu}\bar{\nabla}_{\alpha}\bar{\nabla}_{\beta}+\bar{g}_{\alpha\beta}\bar{\nabla}_{\mu}\bar{\nabla}_{\nu}\right)\nonumber \\
 & +\frac{1}{2}\left(\bar{g}_{\mu\alpha}\bar{\nabla}_{\beta}\bar{\nabla}_{\nu}+\bar{g}_{\nu\beta}\bar{\nabla}_{\alpha}\bar{\nabla}_{\mu}\right).
\end{align}

After this small digression to  Einstein's theory, let us go back to the generic gravity, and split the full Bianchi Identity
\begin{equation}
\nabla_{\mu}\Phi^{\mu\nu}\left(g,R,\nabla R,R^{2},...\right)=0.
\end{equation}
We assume that the theory comes from a diffeomorphism invariant action and this Bianchi Identity is identically satisfied for all smooth metrics. (Note that even if this equation is satisfied only on-shell, we can still construct conserved charges, but at this stage, let us not consider these exotic theories without action, an example can be found in \cite{BTexotic}.)  So we have up to second order 
\begin{equation}
\nabla_{\mu}\Phi^{\mu\nu} = \bar \nabla_\mu \bar \Phi^{\mu \nu} + \kappa \bar \nabla_\mu \Phi^{(1)\mu \nu}  + \kappa \big(\nabla_\mu \big)^{(1)}\bar  \Phi^{\mu \nu}   + {O}(\kappa^2) =0.
\label{linBi}
\end{equation}
Here a somewhat non-trivial-looking piece is the third term which reads 
\begin{equation}
\big(\nabla_\mu \big)^{(1)}\bar  \Phi^{\mu \nu}  = \big ( \Gamma^\mu_{\mu \alpha} \big)^{(1)} \bar  \Phi^{\alpha \nu} +\big ( \Gamma^\nu_{\mu \alpha} \big)^{(1)} \bar  \Phi^{\mu \alpha},
\end{equation}
which,  just like the first term in (\ref{linBi}) vanishes when the background field equation is used. Hence we are left with the linearized Bianchi Identity that will lead us to the conserved charges: 
\begin{tcolorbox}
\begin{equation}
\bar \nabla_\mu \Phi^{(1)\mu \nu} =0.
\label{linbia}
\end{equation}
\end{tcolorbox}
\noindent This is a necessary ingredient in building conserved charges, but not a sufficient one since $\Phi^{(1)\mu \nu}$ is covariantly conserved, not partially conserved: we also need background symmetries.  Of course, from a physical point of view this is what one expects: without symmetries, we really would not know what it means to have conservation. We can define some charges but unless they are conserved, what would they mean?
 So we assume the existence of some (at least one time-like) background Killing vectors forming the set 
$ \{ \bar \xi^\mu_{(a)} \}$  where the index $a$ labels different Killing vectors
\begin{eqnarray}
{\bar{\nabla}}_{\mu}\bar{\xi}_{\nu}^{(a)}+{\bar{\nabla}}_{\nu}\bar{\xi}_{\mu}^{(a)}=0.
\end{eqnarray}
Making use of the linearized Bianchi Identity  (\ref{linbia}) and a given Killing vector, we have a partially conserved current
\begin{equation}
\sqrt{-\bar{g}}\bar{\nabla}_{\mu}\big (\Phi^{(1)\mu \nu}\bar{\xi}_{\nu}^{(a)}\big)=\partial_{\mu}\big(\sqrt{-\bar{g}}\Phi^{(1)\mu \nu}\bar{\xi}_{\nu}^{(a)}\big)=0.
\end{equation}
So we have  $\partial_\mu \mathcal{J}_{(a)}^\mu=0$ where the desired current is
\begin{tcolorbox}
\begin{equation}
\mathcal{J}^\mu := \sqrt{-\bar{g}}\Phi^{(1)\mu \nu}\bar{\xi}_{\nu}.
\label{suslu}
\end{equation}
\end{tcolorbox}
\noindent For notational simplicity we dropped the $a$ index. $\Phi^{(1)\mu \nu}$ is an interesting object: we can directly compute it as the first order term in the expansion of the field equations (\ref{ope-1}). So in that sense, it is a relatively easy linear object to compute, given the field equations. But via the field equations, it is also of the form
\begin{equation}
\Phi^{(1)\mu \nu} = \tau^{\mu \nu} - \sum_{i=1}^\infty \kappa^i  \Phi^{( i +1)\mu \nu} (h),
\end{equation}
that includes all localized matter plus all the non-linear in $h$-terms.  So, far away from the bounded sources,  one still has a non-zero current
\begin{equation}
 \mathcal{J}^\mu \rightarrow  -\sqrt{-\bar g}\bar \xi_\nu \sum_{i=1}^\infty \kappa^i  \Phi^{( i +1)\mu \nu} (h),
\end{equation}
which arises from the non-linearity of the theory and from the fact that gravity itself carries energy.

From the partially conserved $\mathcal{J}^\mu $, it is now easy to construct a global conserved charge. Let $\bar{\mathcal{M}}$ denote the background geometry with zero total charges, and let $h_{\mu \nu}$ denote a symmetric field living on it. Then we have
\begin{equation}
0= \int_{\bar{\mathcal{M}}} d^n x \,\partial_\mu \mathcal{J}^\mu = \int_{\partial {\bar{\mathcal{M}} }} d^{n-1}y\, {\hat n}_\mu \mathcal{J}^\mu, 
\label{equal-zero}
\end{equation}
where we have used the Stokes' theorem; $\partial {\bar{\mathcal{M}} }$ is the total boundary of the spacetime which we assume to be non-null and ${\hat n}_\mu $ is the unit normal co-vector to the boundary (time-like on the spacelike pieces of the boundary and space-like on the time-like piece).   See  Figure 1\footnote{I would like to thank Emel Altas for the figure.}, the side parts of this ``cylindrical'' spacetime constitute a time-like boundary at spatial infinity where we assume that there is no current.
\begin{figure}
\begin{centering}
\includegraphics[scale=0.5]{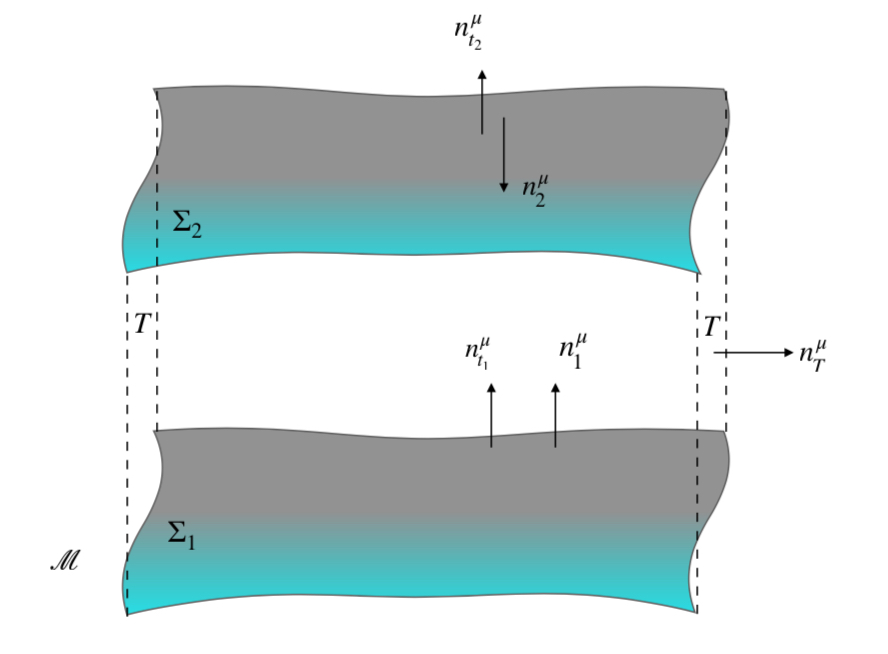} 
\par\end{centering}
\caption{$\Sigma_{1}$ and $\Sigma_{2}$ and $T$ are co-dimension one hypersurfaces over which
the integration is taken. $\Sigma_{1}$ and $\Sigma_{2}$ are constant-time slices, hence they are spacelike, and their normal vectors
$n_{t_{1}}^{\mu}$ and $n_{t_{2}}^{\mu}$, respectively, are time-like. 
The surface normals are taken into the bulk of spacetime in the application of Stokes'
theorem, and these surface normals are represented as $n_{1}^{\mu}$
and $n_{2}^{\mu}$ for two space-like surfaces. The time-like surface $T$ is at spatial infinity and we assume that there is no flux on it.}
\end{figure}

The fact that we get zero when ${\hat n}_\mu \mathcal{J}^\mu$ is integrated all over the boundary is expected, so we define 
integration over a spatial hypersurface $\bar{\Sigma}_1$ which is {\it not equal to the total boundary $\partial {\bar{\mathcal{M}} }$} as
\begin{equation}
Q[\bar \xi, \bar{\Sigma}_1 ]:= \int_{\bar{\Sigma}_1 } d^{n-1}y\, {\hat n}_\mu \mathcal{J}^\mu =\int_{\bar{\Sigma}_1 } d^{n-1}y\
\sqrt{\bar{\gamma}}\, {\hat n}_\mu \bar{\xi}_{\nu}\Phi^{(1)\mu \nu},
\label{charge1}
\end{equation}
where $\gamma_{\mu \nu}$ is the pullback metric on $\bar{\Sigma}_1$.  Now assume that we have $\partial {\bar{\mathcal{M}} } =  \bar{\Sigma}_1 \cup \bar{\Sigma}_2\cup \bar T $,  then the statement (\ref{equal-zero}) can be restated as a charge-conservation statement as
\begin{equation}
Q[\bar \xi, \bar{\Sigma}_1 ] =Q[\bar \xi, \bar{\Sigma}_2 ].
\end{equation} 
So we can drop the reference to which hypersurface we refer to (risking to be pedantic, let us note that any spatial surface at time $t$, $\bar \Sigma_t$ will do the job and yield the same value for the charge, hence the time-independence, or conservation of the charge). For a given theory, we can work with (\ref{charge1}), but this is not advised as there will usually be singularities (of the black hole type) in the bulk, and therefore one should convert this integral to a boundary integral using the Stokes' theorem one more time.  Clearly boundary of a boundary  is empty ($ \partial \partial \bar {\mathcal{M}}  = \emptyset$), but generically $\partial \bar \Sigma_t \ne \emptyset$.  To use Stokes' theorem in (\ref{charge1}), we need one more ingredient: since $\bar{\xi}_{\nu}\Phi^{(1)\mu \nu}$ is identically (i.e. off-shell) divergence-free and is built locally from $h_{\mu\nu}$, the algebraic Poincar\'e lemma \cite{Wald1990} guarantees that there is an antisymmetric potential  $\mathcal{F}^{\mu \nu}(\bar \xi, h)$  such that
\begin{equation}
\bar{\xi}_{\nu}\Phi^{(1)\mu \nu} =: \bar \nabla_\nu  \mathcal{F}^{\mu \nu}(\bar \xi, h).
\label{F_def}
\end{equation}
Of course, at this stage, we keep $\mathcal{F}^{\mu \nu}(\bar \xi, h)$ implicit, and it can be computed once the gravity theory is given.  So then we have 
\begin{equation}
Q[\bar \xi] =\int_{\bar{\Sigma}} d^{n-1}y\
\sqrt{\bar{\gamma}}\, {\hat n}_\mu \bar \nabla_\nu  \mathcal{F}^{\mu \nu}(\bar \xi, h) = \int_{\partial \bar{\Sigma}} d^{n-2}z\
\sqrt{\bar{q}}\, {\hat n}_\mu \hat \sigma_\nu  \mathcal{F}^{\mu \nu}(\bar \xi, h), 
\label{charge2}
\end{equation}
where $\bar q_{\mu \nu}$ is the induced metric on $\partial \bar{\Sigma}$; and  $\hat \sigma_\nu$ is the outward unit normal co-vector on it.   The second equality is most easily seen in coordinates adapted to the slicing, $\bar\Sigma=\{t=\text{const}\}$: for an antisymmetric tensor one has $\sqrt{-\bar g}\,\bar\nabla_\nu \mathcal{F}^{\mu\nu}=\partial_\nu\big(\sqrt{-\bar g}\,\mathcal{F}^{\mu\nu}\big)$, and $\sqrt{\bar\gamma}\,\hat n_\mu V^\mu=-\sqrt{-\bar g}\,V^0$ for any vector $V^\mu$, so that the bulk integral becomes $-\int_{\bar\Sigma} d^{n-1}x\,\partial_i\big(\sqrt{-\bar g}\,\mathcal{F}^{0i}\big)$, which is a surface integral over $\partial\bar\Sigma$ (since $\mathcal{F}^{00}=0$). Equivalently, $Q$ is the integral of the $(n-2)$-form dual to $\mathcal{F}$ over $\partial\bar\Sigma$.

We can summarize the result in a more (anti)-symmetric way using the binormal vector $\bar{\epsilon}_{\mu \nu} $ as
\begin{tcolorbox}
 \begin{equation}
Q[\bar \xi] = \int_{\partial \bar{\Sigma}} d^{n-2}z\
\sqrt{\bar{q}}\, \bar{\epsilon}_{\mu \nu} \mathcal{F}^{\mu \nu}(\bar \xi, h),  \hskip 1 cm \bar{\epsilon}_{\mu \nu} :=\frac{1}{2}\left (\hat n_\mu \hat \sigma_\nu -{\hat n}_\nu \hat \sigma_\mu \right ).
\label{charge3}
\end{equation}
\end{tcolorbox}
\noindent This is a generic formula valid for all gravity theories: one just needs to compute the antisymmetric tensor $\mathcal{F}^{\mu \nu}(\bar \xi, h)$ of the relevant theory and the background quantities $\bar \xi^\mu$,  $\bar q_{\mu \nu}$, and  $\bar{\epsilon}_{\mu \nu}$.

Let us remark on a couple of points:

\begin{enumerate}
\item Given the gravity theory,  explicit construction of $\mathcal{F}^{\mu \nu}(\bar \xi, h)$ is by no means a trivial task; also from its very definition (\ref{F_def}) it is not unique:  $\mathcal{F}^{\mu \nu}(\bar \xi, h)$ and  $\mathcal{F}^{\mu \nu}(\bar \xi, h)+ \Upsilon^{\mu \nu }(\bar \xi, h)$ give the same total charges as long as $\nabla_\nu \Upsilon^{\mu \nu }(\bar \xi, h)=0$ and $\Upsilon^{\mu \nu }$ is antisymmetric. 

\item Any astute reader who is well-versed in integration on generic manifolds must have realized that what we have done above would be done in a much more elegant way with differential forms which are the natural objects to be used in integration.  For such a construction see \cite{Benn} where it was shown that mass and angular momenta appear as de Rham periods of closed two forms (in 4-dimensional gravity).  

\end{enumerate}

\subsection{Einstein's Theory with a Cosmological Constant }

To fix the notation properly and make this section self-consistent, let us start with the action
\begin{equation}
I = \frac{1 }{2\kappa} \int_{\mathcal{M}} d^n x\, \sqrt{-g} \left ( R- 2 \Lambda \right) +  I_{\mbox{matter}}[\psi, g^{\mu \nu}],
\label{act1}
\end{equation}
whose gravity-sector field equations are
\begin{equation}
G_{\mu \nu} = R_{\mu \nu} - \frac{1}{2} g_{\mu \nu} R + \Lambda g_{\mu \nu} = \kappa \tau_{\mu \nu}.
\label{fieldeq}
\end{equation}
At this stage, we assume minimal coupling and a localized matter source in a compact region of spacetime.  Note that in $n$ dimensions, one defines the Newton constant via $ \kappa = :2 \Omega_{n-2} G_n$ with $\Omega_{n-2}$ being the solid angle.  For a well-defined variational principle in this theory, we should add a boundary term to the action (\ref{act1}) (see the remark on the Regge-Teitelboim boundary term in the Conclusions), but at this stage, all we need is the field equations  (\ref{fieldeq}) as we have seen in the previous section. Note that the action (\ref{act1}) is normalized with $1/(2\kappa)$, while the higher curvature actions below are normalized with $1/\kappa$; in both cases the matter source is normalized so that the Einstein limit of the field equations reads $G_{\mu\nu}=\kappa\tau_{\mu\nu}$, so $\kappa$ has the same meaning throughout.

To be as general as possible, let us consider an {\it Einstein spacetime } (not necessarily maximally symmetric) as the background solution with 
\begin{equation}
\bar R_{\mu \nu} =  \frac{ 2 \Lambda}{n-2}\bar g_{\mu \nu}, \hskip 2 cm \bar R =  \frac{ 2n \Lambda}{n-2}.
\end{equation}
We require that this Einstein spacetime has a Killing vector, otherwise, we cannot make sense of the conserved charge construction given above.  The first order of business is to compute the linearized Einstein tensor in this background as demanded by ({\ref{F_def}).  From (\ref{fieldeq}) we have the generic form of the linearized Einstein tensor in the background of an Einstein spacetime:
\begin{equation}
\mathcal{G}^{(1)}_{\mu \nu} = R^{(1)}_{\mu \nu} - \frac{1}{2} \bar g_{\mu \nu} R^{(1)}  -\frac{1}{2} h_{\mu \nu} \bar R + \Lambda h_{\mu \nu}.
\label{linEin}
\end{equation}
We give the linearizations of various tensors in generic backgrounds in Appendix A.  Here I shall use them. The linearized Ricci tensor is
\begin{equation}
 R^{(1)}_{\mu \nu} =\frac{1}{2}\left(-{\bar{\Box}}h_{\mu\nu}-{\bar{\nabla}}_{\mu}{\bar{\nabla}}_{\nu}h+{\bar{\nabla}}^{\sigma}{\bar{\nabla}}_{\nu}h_{\sigma\mu}+{\bar{\nabla}}^{\sigma}{\bar{\nabla}}_{\mu}h_{\sigma\nu}\right),\label{linearricci1}
\end{equation}
and the linearized Ricci scalar is
\begin{equation}
R^{(1)} = \bar g^{\mu\nu}R^{(1)}_{\mu \nu}  - \bar R_{\mu\nu} h^{\mu \nu}.
\end{equation}
Therefore using the last two equations in (\ref{linEin}) we have
\begin{eqnarray}
\mathcal{G}^{(1)}_{\mu \nu} =&& \frac{1}{2}\left(-{\bar{\Box}}h_{\mu\nu}-{\bar{\nabla}}_{\mu}{\bar{\nabla}}_{\nu}h+{\bar{\nabla}}^{\sigma}{\bar{\nabla}}_{\nu}h_{\sigma\mu}+{\bar{\nabla}}^{\sigma}{\bar{\nabla}}_{\mu}h_{\sigma\nu}\right) -\frac{1}{2} h_{\mu \nu} \bar R + \Lambda h_{\mu \nu} \nonumber \\
&&-\frac{1}{2}\bar g_{\mu\nu} \left(-\bar{\Box}h+\bar{\nabla}_{\rho}\bar{\nabla}_{\sigma}h^{\rho\sigma}- \bar 
R_{\rho\sigma}h^{\rho \sigma}\right).
\end{eqnarray}
To proceed, and learn some lessons for higher derivative theories, let us rewrite the last equation as derivatives plus non-derivative terms.
\begin{eqnarray}
\mathcal{G}^{(1)\mu \nu}  = &&\frac{1}{2} \bar \nabla_\alpha \bar \nabla_\beta  \Big ( \bar g^{\mu \beta} h^{\nu \alpha} +\bar g^{\nu \beta} h^{\mu \alpha} -\bar g^{\alpha \beta} h^{\mu \nu} -\bar g^{\mu \nu} h^{ \alpha\beta} +\bar g^{\mu \nu } \bar g^{ \alpha \beta} h -\bar g^{\mu \alpha} \bar g^{ \nu \beta} h  \Big ) \nonumber \\
&& +\frac{1}{2} \bar g^{\mu \nu} \bar 
R_{\rho\sigma}h^{\rho \sigma} -\frac{1}{2} h^{\mu \nu} \bar R + \Lambda h^{\mu \nu}. 
\label{symmetry2}
\end{eqnarray} 
We want to rewrite this as follows\footnote{We do not require separate symmetry in the $\mu$ and $\nu$ indices for the derivative and non-derivative terms, but together, they are of course symmetric.}
\begin{equation}
\mathcal{G}^{(1)\mu \nu}  =: \bar \nabla_\alpha \bar \nabla_\beta  K^{\mu \alpha \nu \beta} + X^{\mu \nu},
\end{equation}
where $K^{\mu \alpha \nu \beta}$ has the symmetries of the Riemann tensor and $X^{\mu \nu}$, the ``fudge tensor'', should  hopefully vanish at the end. Rearranging the first line in (\ref{symmetry2}), one arrives at the Abbott-Deser superpotential given in (\ref{sup})
\begin{eqnarray}
K^{\mu\alpha\nu\beta}:=\frac{1}{2}\Big (\bar{g}^{\mu\beta}\tilde h^{\nu\alpha}+\bar{g}^{\nu\alpha}\tilde h^{\mu\beta}-\bar{g}^{\mu\nu}\tilde h^{\alpha\beta}-\bar{g}^{\alpha\beta}\tilde h^{\mu\nu}\Big ),\hskip0.7 cm \tilde h^{\mu\nu}:=h^{\mu\nu}-\frac{1}{2}\bar{g}^{\mu\nu}h,
\label{sup2}
\end{eqnarray}
while the fudge tensor reads
\begin{eqnarray}
X^{\mu \nu} :&=& \frac{1}{2} \bar g^{\nu \beta} [ \bar \nabla_\alpha, \bar \nabla_\beta ]h^{\mu \alpha} + \frac{1}{2} \bar g^{\mu \nu} \bar R_{\rho\sigma}h^{\rho \sigma} -\frac{1}{2} h^{\mu \nu} \bar R + \Lambda h^{\mu \nu}\nonumber \\
& =&\frac{1}{2} \big (\bar R^\nu_\alpha h^{\mu \alpha}- \bar R^{\mu \alpha \nu \beta}h_{\alpha \beta} \big) + \frac{1}{2} \bar g^{\mu \nu} \bar R_{\rho\sigma}h^{\rho \sigma} -\frac{1}{2} h^{\mu \nu} \bar R + \Lambda h^{\mu \nu}.
\end{eqnarray}
Finally, we have\footnote{We use the identity $ \bar \nabla_\beta\bar \nabla_\alpha \bar \xi_\nu = \bar R^\rho\,_{\beta \alpha \nu} \bar \xi_\rho$ which is valid for a Killing vector on a (pseudo-)Riemannian manifold.} 
\begin{eqnarray}
\bar \xi_\nu \mathcal{G}^{(1)\mu \nu}  &=&\bar \xi_\nu \bar \nabla_\alpha \bar \nabla_\beta  K^{\mu \alpha \nu \beta} +\bar \xi_\nu X^{\mu \nu} \nonumber \\
&=&\bar \nabla_\alpha \big (  \bar \xi_\nu \bar \nabla_\beta  K^{\mu \alpha \nu \beta} - K^{\mu \beta \nu \alpha} \bar \nabla_\beta \bar \xi_\nu\big )  + K^{\mu \alpha \nu \beta}\bar R_{\rho \beta \alpha \nu} \bar \xi^\rho +\bar \xi_\nu X^{\mu \nu} .
\end{eqnarray}
We found the anti-symmetric background tensor $\mathcal{F}^{\mu \alpha}$   as the term in the bracket, but the additional two terms must somehow vanish.  It is easy to show that the additional two terms can be written as 
\begin{equation}
K^{\mu \alpha \nu \beta}\bar R_{\rho \beta \alpha \nu} \bar \xi^\rho +\bar \xi_\nu X^{\mu \nu}  = \bar G_{\rho \sigma} \Big ( \bar\xi^\sigma h^{\rho \mu} +\frac{1}{2} \bar \xi^\mu h^{\sigma \rho}- \frac{1}{2} \bar \xi^\rho \bar g^{\mu \sigma} h  \Big ),
\end{equation}
which vanishes for Einstein backgrounds, $\bar G_{\rho \sigma}=0$, and therefore,  we have 

\begin{eqnarray}
\bar \xi_\nu \mathcal{G}^{(1)\mu \nu}  =\bar \nabla_\alpha \big (  \bar \xi_\nu \bar \nabla_\beta  K^{\mu \alpha \nu \beta} - K^{\mu \beta \nu \alpha} \bar \nabla_\beta \bar \xi_\nu\big )  =: \bar \nabla_\alpha \mathcal{F}^{\mu \alpha}.
\label{denks}
\end{eqnarray}
Using (\ref{charge3}), we get the desired result:
\begin{tcolorbox}
 \begin{equation}
Q[\bar \xi] = \frac{1}{2 \Omega_{n-2} G_n}\int_{\partial \bar{\Sigma}} d^{n-2}z\
\sqrt{\bar{q}}\, \bar{\epsilon}_{\mu \alpha}\big (  \bar \xi_\nu \bar \nabla_\beta  K^{\mu \alpha \nu \beta} - K^{\mu \beta \nu \alpha} \bar \nabla_\beta \bar \xi_\nu\big ),  \hskip 0.2 cm \bar{\epsilon}_{\mu \nu} :=\frac{1}{2}\left (\hat n_\mu \hat \sigma_\nu -{\hat n}_\nu \hat \sigma_\mu \right ),
\label{charge4}
\end{equation}
\end{tcolorbox}
\noindent where we have normalized the charges with a constant overall factor. Let us recall all the ingredients in this expression and the region of its applicability.  First, we assume that spacetime allows the following splitting 
$\bar {\mathcal{M}}= \mathbb{R} \times \bar \Sigma$ and we assume that a generic Einstein metric $\bar g_{\mu \nu}$ solves the Einstein's equation without a source on this spacetime and $\bar g_{\mu \nu}$  has a Killing symmetry $\xi$; and the total boundary of spacetime  $\partial \bar {\mathcal{M}}$ is non-empty.  Then in any coordinates that assign a zero conserved charge to the background metric, the above formula gives the conserved charge of the deviated geometry on the same manifold with the metric $ g_{\mu \nu} = \bar g_{\mu \nu} + \kappa h_{\mu \nu}$. The formula has the following unifying features: it gives the ADM energy, momentum, and angular momentum for flat backgrounds, it gives the same conserved quantities for (Anti)-de Sitter backgrounds; moreover, it generalizes to the generic Einstein backgrounds that possess a Killing symmetry.  It is also refreshing that one does not have to work in particular coordinates, we have not chosen gauge in the derivation so far.

Let us give another, but equivalent, expression $\bar{\xi}_{\nu}{\cal G}_{(1)}^{\mu\nu}$ given in \cite{4}. Let us rewrite (\ref{denks}) 

\begin{equation}
\bar{\xi}_{\nu}{\cal G}_{(1)}^{\mu\nu}=\bar{\nabla}_{\alpha}\left(\bar{\xi}_{\nu}\bar{\nabla}_{\beta}K^{\mu\alpha\nu\beta}-K^{\mu\beta\nu\alpha}\bar{\nabla}_{\beta}\bar{\xi}_{\nu}\right).
\end{equation}

Then, the first term has the form
\begin{align}
\bar{\xi}_{\nu}\bar{\nabla}_{\beta}K^{\mu\alpha\nu\beta}
= & \frac{1}{2}\bar{\xi}_{\nu}\bar{\nabla}_{\beta}\left(\bar{g}^{\alpha\nu}h^{\mu\beta}+\bar{g}^{\mu\beta}h^{\alpha\nu}-\bar{g}^{\alpha\beta}h^{\mu\nu}-\bar{g}^{\mu\nu}h^{\alpha\beta}+\bar{g}^{\mu\nu}\bar{g}^{\alpha\beta}h-\bar{g}^{\alpha\nu}\bar{g}^{\mu\beta}h\right)\nonumber \\
= & \frac{1}{2}\left(\bar{\xi}^{\alpha}\bar{\nabla}_{\beta}h^{\mu\beta}-\bar{\xi}^{\mu}\bar{\nabla}_{\beta}h^{\alpha\beta}+\bar{\xi}_{\nu}\bar{\nabla}^{\mu}h^{\alpha\nu}-\bar{\xi}_{\nu}\bar{\nabla}^{\alpha}h^{\mu\nu}+\bar{\xi}^{\mu}\bar{\nabla}^{\alpha}h-\bar{\xi}^{\alpha}\bar{\nabla}^{\mu}h\right),
\end{align}
and the second term becomes
\begin{align}
K^{\mu\beta\nu\alpha}\bar{\nabla}_{\beta}\bar{\xi}_{\nu}=  & \frac{1}{2}\left(-h^{\mu\nu}\bar{\nabla}^{\alpha}\bar{\xi}_{\nu}+h^{\alpha\nu}\bar{\nabla}^{\mu}\bar{\xi}_{\nu}+h\bar{\nabla}^{\alpha}\bar{\xi}^{\mu}\right).
\end{align}
Thus, one has
\begin{align}
2\bar{\xi}_{\nu}{\cal G}_{(1)}^{\mu\nu}=\bar{\nabla}_{\alpha}\biggl( & \bar{\xi}^{\alpha}\bar{\nabla}_{\beta}h^{\mu\beta}-\bar{\xi}^{\mu}\bar{\nabla}_{\beta}h^{\alpha\beta}+\bar{\xi}_{\nu}\bar{\nabla}^{\mu}h^{\alpha\nu}-\bar{\xi}_{\nu}\bar{\nabla}^{\alpha}h^{\mu\nu}+\bar{\xi}^{\mu}\bar{\nabla}^{\alpha}h-\bar{\xi}^{\alpha}\bar{\nabla}^{\mu}h\nonumber \\
 & +h^{\mu\nu}\bar{\nabla}^{\alpha}\bar{\xi}_{\nu}-h^{\alpha\nu}\bar{\nabla}^{\mu}\bar{\xi}_{\nu}-h\bar{\nabla}^{\alpha}\bar{\xi}^{\mu}\biggr),
\end{align}
which is the same result as the one given in  \cite{4}. So in this form, the conserved charges read as
\begin{tcolorbox}
\begin{eqnarray}\label{adt1}
Q^{\mu}(\bar{\xi})=\frac{1}{4\Omega_{n-2}G_n}\int_{\partial \bar{\Sigma}} & dS_{i} & \Big\{\bar{\xi}_{\nu}\bar{\nabla}^{\mu}h^{i\nu}-\bar{\xi}_{\nu}\bar{\nabla}^{i}h^{\mu\nu}+\bar{\xi}^{\mu}\bar{\nabla}^{i}h-\bar{\xi}^{i}\bar{\nabla}^{\mu}h\\
 &  & +h^{\mu\nu}\bar{\nabla}^{i}\bar{\xi}_{\nu}-h^{i\nu}\bar{\nabla}^{\mu}\bar{\xi}_{\nu}+\bar{\xi}^{i}\bar{\nabla}_{\nu}h^{\mu\nu}-\bar{\xi}^{\mu}\bar{\nabla}_{\nu}h^{i\nu}+h\bar{\nabla}^{\mu}\bar{\xi}^{i}\Big\} ,\nonumber
\end{eqnarray}
\end{tcolorbox}
and $\mu=0$ should be taken.

\subsection{Gauge invariance issue}

The total conserved charges we have found should certainly be invariant under small gauge transformations. Let us discuss this issue in some detail.

Under small diffeomorphism generated by an infinitesimal
arbitrary vector field $\zeta^{\mu}$, one has 
\begin{equation}
\delta_{\zeta}h_{\mu\nu}=\mathcal{L}_{\zeta}g_{\mu\nu}=\nabla_{\mu}\zeta_{\nu}+\nabla_{\nu}\zeta_{\mu},
\end{equation}
and if one considers the first order in the small parameters $h_{\mu\nu}$
and $\zeta^{\mu}$, the first order of the transformation becomes
\begin{equation}
\delta_{\zeta}h_{\mu\nu}=\bar{\nabla}_{\mu}\zeta_{\nu}+\bar{\nabla}_{\nu}\zeta_{\mu}.
\end{equation}
Any scalar field $\phi$ transforms as 
\begin{equation}
\delta_{\zeta}\phi=\mathcal{L}_{\zeta}\phi=\zeta^{\mu}\partial_{\mu}\phi,
\end{equation}
then, the scalar
curvature has the transformation
\begin{equation}
\delta_{\zeta}R=\zeta^{\mu}\partial_{\mu}R,
\end{equation}
that is
\begin{equation}
R^{\prime}(x)=R(x)+\zeta^{\mu}(x)\partial_{\mu}R(x).
\end{equation}
Then, the linearized scalar curvature $R_{L}$ transforms as
\begin{equation}
\bar{R}+R_{L}^{\prime}+O\left(h^{\prime2}\right)=\bar{R}+R_{L}+\zeta^{\mu}\partial_{\mu}\bar{R}+O\left(h^{2}\right)+O\left(\zeta h\right),
\end{equation}
then we have a gauge invariant linearized scalar curvature:
\begin{equation}
\delta_{\zeta}R_{L}=\zeta^{\mu}\partial_{\mu}\bar{R} =0
\end{equation}
for Einstein spaces since $\bar{R}$ is constant. 
Similarly the Ricci tensor $R_{\mu\nu}$ transforms as 
\begin{equation}
\delta_{\zeta}R_{\mu\nu}=\zeta^{\rho}\nabla_{\rho}R_{\mu\nu}+\left(\nabla_{\mu}\zeta^{\rho}\right)R_{\rho\nu}+\left(\nabla_{\nu}\zeta^{\rho}\right)R_{\mu\rho},
\end{equation}
and $R_{\mu\nu}^{(1)}$ changes as
\begin{equation}
R_{\mu\nu}^{\prime(1)}=R_{\mu\nu}^{(1)}+\zeta^{\rho}\bar{\nabla}_{\rho}\bar{R}_{\mu\nu}+\left(\bar{\nabla}_{\mu}\zeta^{\rho}\right)\bar{R}_{\rho\nu}+\left(\bar{\nabla}_{\nu}\zeta^{\rho}\right)\bar{R}_{\mu\rho},
\end{equation}
for $\bar{R}_{\mu\nu}=\frac{2\Lambda}{n-2}\bar{g}_{\mu\nu}$, we have 
\begin{align}
R_{\mu\nu}^{\prime{(1)}} & =R_{\mu\nu}^{{(1)}}+\frac{2\Lambda}{n-2}\left(\bar{\nabla}_{\mu}\zeta_{\nu}+\bar{\nabla}_{\nu}\zeta_{\mu}\right)
  =R_{\mu\nu}^{{(1)}}+\frac{2\Lambda}{n-2}\delta_{\zeta}h_{\mu\nu}.
\end{align}
Finally, for $G_{\mu\nu}=R_{\mu\nu}-\frac{1}{2}g_{\mu\nu}R+\Lambda g_{\mu\nu}$,
one has
\begin{equation}
\delta_{\zeta}G_{\mu\nu}=\zeta^{\rho}\nabla_{\rho}G_{\mu\nu}+\left(\nabla_{\mu}\zeta^{\rho}\right)G_{\rho\nu}+\left(\nabla_{\nu}\zeta^{\rho}\right)G_{\mu\rho},
\end{equation}
 and the linearized Einstein tensor transforms as
\begin{equation}
\delta_{\zeta}\mathcal{G}^{(1)}_{\mu\nu}=\zeta^{\rho}\bar{\nabla}_{\rho}\bar{G}_{\mu\nu}+\left(\bar{\nabla}_{\mu}\zeta^{\rho}\right)\bar{G}_{\rho\nu}+\left(\bar{\nabla}_{\nu}\zeta^{\rho}\right)\bar{G}_{\mu\rho} =0,
\end{equation}
where the last equation follows from the fact that we are working in Einstein spaces.

The above result gives what we want: $\delta_{\zeta} ( \bar{\xi}_{\nu}{\cal G}_{(1)}^{\mu\nu}) =0$ and hence the total charges are gauge invariant. 
On the other hand, the anti-symmetric tensor ${\cal {F}}^{\alpha\mu}$ that appears as
\begin{equation}
\bar{\xi}_{\nu}({\cal G}^{\mu\nu})^{\left(1\right)}=\bar{\nabla}_{\alpha}{\cal {F}}^{\alpha\mu},\label{equationofj}
\end{equation}
which reads explicitly as
\begin{align}
2{\cal {F}}^{\alpha\mu}:= & \bar{\xi}^{\alpha}\bar{\nabla}_{\beta}h^{\mu\beta}-\bar{\xi}^{\mu}\bar{\nabla}_{\beta}h^{\alpha\beta}+\bar{\xi}_{\nu}\bar{\nabla}^{\mu}h^{\alpha\nu}-\bar{\xi}_{\nu}\bar{\nabla}^{\alpha}h^{\mu\nu}+\bar{\xi}^{\mu}\bar{\nabla}^{\alpha}h-\bar{\xi}^{\alpha}\bar{\nabla}^{\mu}h\label{explicitjtensor} \nonumber \\
 & +h^{\mu\nu}\bar{\nabla}^{\alpha}\bar{\xi}_{\nu}-h^{\alpha\nu}\bar{\nabla}^{\mu}\bar{\xi}_{\nu}-h\bar{\nabla}^{\alpha}\bar{\xi}^{\mu},
\end{align}
is {\it not} gauge invariant. In fact, under gauge transformations, it transforms as a boundary term that does not affect the total conserved charge. 
\begin{equation}
2 \delta_{\zeta}{\cal {F}}^{\alpha\mu}=\bar{\nabla}_{\nu}\left(\bar{\xi}^{\alpha}\bar{\nabla}^{\nu}\zeta^{\mu}+\bar{\xi}^{\mu}\bar{\nabla}^{\alpha}\zeta^{\nu}+\bar{\xi}^{\nu}\bar{\nabla}^{\mu}\zeta^{\alpha}+2\zeta^{\alpha}\bar{\nabla}^{\nu}\bar{\xi}^{\mu}+\zeta^{\nu}\bar{\nabla}^{\mu}\bar{\xi}^{\alpha}-(\mu\leftrightarrow\alpha)\right).
\end{equation}
One might wonder if an explicitly gauge invariant ${\cal {F}}^{\alpha\mu}$ can be constructed. This turns out to be possible for asymptotically (A)dS spacetime and the final expression for the total charges is very satisfying. We shall now describe this briefly and leave the details to the original works \cite{emel1,emel2}.
For this purpose, one needs the following rank 4 tensor  that has the algebraic symmetries of the Riemann tensor
\begin{equation}
\text{\ensuremath{{\cal {P}}}}^{\nu\mu\beta\sigma}:=R^{\nu\mu\beta\sigma}+g^{\sigma\nu}R^{\beta\mu}-g^{\beta\nu}R^{\sigma\mu}+g^{\beta\mu}R^{\sigma\nu}-g^{\sigma\mu}R^{\beta\nu}+\left(\frac{R}{2}-\frac{\Lambda\left(n-3\right)}{n-1}\right)\left(g^{\beta\nu}g^{\sigma\mu}-g^{\sigma\nu}g^{\beta\mu}\right),\label{eq:Ptensor}
\end{equation}
which is divergence-free without the use of any field equation but just the Bianchi Identities
\begin{equation}
\nabla_{\nu}{\cal {P}}^{\nu}\thinspace_{\mu\beta\sigma}=0.\label{divP}
\end{equation}
Moreover, its trace is the cosmological Einstein tensor: 
\begin{equation}
{\cal {P}}^{\nu}\thinspace_{\mu\nu\sigma}=(3-n)\text{\ensuremath{{\cal {G}}}}_{\mu\sigma}.
\end{equation}
This last property was our motivation to construct the ${\cal {P}}$ tensor in \cite{emel1} but its divergence-free property turned out to be very useful for the conserved charge construction in an explicitly gauge invariant way.\footnote{The same construction without linearization gives a nice expression for the surface gravity and the Hawking temperature of black holes in terms of an integral of the Gauss-Bonnet invariant outside the black hole region of spacetime \cite{emelhawk}.}  This works as follows: for any anti-symmetric tensor $\cal {X}_{\beta\sigma}$, we have the exact identity

\begin{equation}
\nabla_{\nu}(\text{\ensuremath{{\cal {X}}}}_{\beta\sigma}\text{\ensuremath{{\cal {P}}}}^{\nu\mu\beta\sigma})-\text{\ensuremath{{\cal {P}}}}^{\nu\mu\beta\sigma}\nabla_{\nu}\text{\ensuremath{{\cal {X}}}}_{\beta\sigma}=0.\label{eq:ddimensionalmainequation}
\end{equation}
Linearization of this relation about an $AdS$ background gives 
\begin{equation}
\bar{\nabla_{\nu}}\biggl((\text{\ensuremath{{\cal {P}}}}^{\nu\mu\beta\sigma})^{\left(1\right)}\bar{\text{\ensuremath{{\cal {X}}}}}_{\beta\sigma}\biggr)-(\text{\ensuremath{{\cal {P}}}}^{\nu\mu\beta\sigma})^{\left(1\right)}\bar{\nabla}_{\nu}\bar{\text{\ensuremath{{\cal {X}}}}}_{\beta\sigma}=0,\label{eq:ddimensionalmainequationlinear}
\end{equation}
where the linearized ${\cal {P}}$ tensor reads
\begin{eqnarray}
({\cal {P}}^{\nu\mu\beta\sigma})^{\left(1\right)}= &  & (R^{\nu\mu\beta\sigma})^{(1)}+2(\text{\ensuremath{{\cal {G}}}}^{\mu[\beta})^{(1)}\overline{g}^{\sigma]\nu}+2(\text{\ensuremath{{\cal {G}}}}^{\nu[\sigma})^{(1)}\overline{g}^{\beta]\mu}+(R)^{\left(1\right)}\overline{g}^{\mu[\beta}\overline{g}^{\sigma]\nu}\nonumber \\
 &  & +\frac{4\Lambda}{(n-1)(n-2)}(h^{\mu[\sigma}\overline{g}^{\beta]\nu}+\overline{g}^{\mu[\sigma}{h}^{\beta]\nu}),\label{ktensorlinear}
\end{eqnarray}
with square brackets denoting antisymmetrization with a factor
of 1/2. Now at this stage, we make a crucial choice of our anti-symmetric tensor: we take it to be the ``field strength'' of the Killing vector as 
\begin{equation}
\bar{\text{\ensuremath{{\cal {X}}}}}_{\alpha\beta}:=\bar{\nabla}_{\alpha}\bar{\xi}_{\beta},
\end{equation}
which then reduces  (\ref{eq:ddimensionalmainequationlinear}) to
\begin{equation}
\bar{\text{\ensuremath{\xi}}}_{\lambda}(\text{\ensuremath{{\cal {G}}}}^{\lambda\mu})^{\left(1\right)}=\frac{(n-1)(n-2)}{4\Lambda\left(n-3\right)}\bar{\nabla_{\nu}}\biggl((\text{\ensuremath{{\cal {P}}}}^{\nu\mu\beta\sigma})^{\left(1\right)}\bar{\text{\ensuremath{{\cal {X}}}}}_{\beta\sigma}\biggr).\label{eq:finallinearequation}
\end{equation}
So in this new formulation, the anti-symmetric tensor is 
\begin{equation}
{\cal {F}}^{\nu\mu} := \frac{(n-1)(n-2)}{4\Lambda\left(n-3\right)}\biggl((\text{\ensuremath{{\cal {P}}}}^{\nu\mu\beta\sigma})^{\left(1\right)}\bar{\text{\ensuremath{{\cal {X}}}}}_{\beta\sigma}\biggr),
\end{equation}
Since, at spatial infinity, the linearized ${\cal P}$ tensor reduces to the linearized Riemann tensor, this yields the total charge as 
\begin{tcolorbox}
\begin{equation}
\phantom{\frac{\frac{\xi}{\xi}}{\frac{\xi}{\xi}}}Q\left(\bar{\xi}\right)=\frac{(n-1)(n-2)}{8(n-3)\Lambda G\Omega_{n-2}}\int_{\partial\bar{\Sigma}}d^{n-2}x\,\sqrt{\bar{\gamma}}\,\bar{\epsilon}_{\mu\nu}\left(R^{\nu\mu}\thinspace_{\beta\sigma}\right)^{\left(1\right)}\bar{\text{\ensuremath{{\cal {X}}}}}^{\beta\sigma}.\phantom{\frac{\frac{\xi}{\xi}}{\frac{\xi}{\xi}}}\label{newcharge2}
\end{equation}
\end{tcolorbox}
\noindent It is easy to see that the linearized Riemann tensor
\begin{equation}
(R^{\nu}\thinspace_{\alpha\beta\sigma})^{\left(1\right)}=\bar{\nabla}_{\beta}(\Gamma_{\sigma\alpha}^{\nu})^{\left(1\right)}-\bar{\nabla}_{\sigma}(\Gamma_{\beta\alpha}^{\nu})^{\left(1\right)},\label{eq:firstorderriemann}
\end{equation}
 is gauge invariant in AdS.  One has 
\begin{equation}
\delta_{\zeta}(R^{\nu\mu}\thinspace_{\beta\sigma})^{\left(1\right)}=\bar{g}^{\alpha\mu}\delta_{\zeta}(R^{\nu}\thinspace_{\alpha\beta\sigma})^{\left(1\right)}-\bar{R}^{\nu}\thinspace_{\alpha\beta\sigma}\delta_{\zeta}h^{\alpha\mu}.\label{gaugeinvarianceriemannuudd}
\end{equation}
Linearized Christoffel connection transforms as
\begin{equation}
\delta_{\zeta}(\Gamma_{\sigma\alpha}^{\nu})^{\left(1\right)}=\bar{\nabla}_{\sigma}\bar{\nabla}_{\alpha}\zeta^{\nu}+\bar{R}^{\nu}\thinspace_{\alpha\rho\sigma}\zeta^{\rho}.
\end{equation}
Therefore one has 
\begin{equation}
\delta_{\zeta}(R^{\nu\mu}\thinspace_{\beta\sigma})^{\left(1\right)}=\text{\ensuremath{\mathscr{L}}}_{\zeta}\bar{R}^{\nu\mu}\thinspace_{\beta\sigma} =0,
\end{equation}
and hence the integrand in (\ref{newcharge2}) is gauge invariant. In the Appendix, we give a rather compact but slightly more mathematical discussion of the gauge invariance issue as given in \cite{Taub}.

Next we apply  (\ref{adt1}) and (\ref{newcharge2}) to several examples.

\subsubsection{Energy of the Schwarzschild black hole in AdS spacetime}

Let us start with a simple example of the $n$-dimensional Schwarzschild AdS metric that solves Einstein's theory (as well as many other theories) with a cosmological constant. In the usual static coordinates, we have for $n \ge 4$ : 
\begin{align}
ds^{2}=-\Big(1-(\frac{r_{0}}{r})^{n-3}-\frac{r^{2}}{l^{2}}\Big) \, dt^{2}+\Big(1-(\frac{r_{0}}{r})^{n-3}-\frac{r^{2}}{l^{2}}\Big)^{-1}\, dr^{2}+r^{2}d\Omega_{n-2}^{2},\label{sds}
\end{align}
 where $l^{2}\equiv{(n-2)(n-1)}/{2\Lambda}$ (so that $l^2<0$ for AdS). The background metric is obtained for $r_0=0$. We define the charges of the background spacetime to be zero.  
 The  Killing vector $\bar \xi^{\mu}= -(\partial_t)^\mu =  (-1,{\bf {0}})$, is time-like for AdS ($\Lambda <0$)  everywhere but only so inside the cosmological horizon for dS ($\Lambda >0$) since one has  $\bar{g}_{\mu\nu}{\bar{\xi}}^{\mu}\bar{\xi}^{\nu}=-(1-{r^{2}}/{l^{2}})$. So from (\ref{adt1}), we have 
\begin{align}
E \equiv Q[\partial_t]= &\frac{1}{4 \Omega_{n-2} G_N}\int_{S_r^{n-2}} r^{n-2} d\Omega_{n-2} \nonumber \\
\times& \Bigg (
 -\bar g_{00} \bar{\nabla}^{0}h^{r 0} +\bar g_{00} \bar{\nabla}^{r}h^{00} - \bar{\nabla}^{r}h +
h^{0 \nu}\bar{\nabla}^r \bar{\xi}_\nu - h^{r \nu}\bar{\nabla}^0
\bar{\xi}_\nu  + \bar{\nabla}_{\nu}h^{r \nu} + h\,\bar{\nabla}^{0}\bar{\xi}^{r} \Bigg).
\label{yawru}
\end{align}
To understand the picture, let us look at the result in 4 dimensions at some finite $r$ coordinate:
\begin{align}
E(r)=\frac{r_{0}}{2G}\frac{1-\frac{r^{2}}{l^{2}}}{1-\frac{r_{0}}{r}-\frac{r^{2}}{l^{2}}}.
\label{finiter}
\end{align}
Naively if we let $r \rightarrow \infty$, we get  $E =  {r_0}/{ 2 G} \equiv M$. But this is legitimate only for AdS and flat spacetimes, on the other hand, for dS spacetime one must consider small black holes that do not affect the location of the cosmological horizon: namely, we must be able to approximate  $1-{r_0}/{r}-{r^2}/{l^2} \approx 1-{r^2}/{l^2}$. Otherwise, as expected we cannot assign a meaningful mass to the black hole in dS spacetimes. Of course, there is no shame in that: without a global time-like coordinate outside a compact region, this is the best one can do. 
  
Generically in $n$ dimensions, the result of (\ref{yawru}) as $r \rightarrow \infty$ is 
\begin{align}
E=\frac{n-2}{4G_{N}}r_{0}^{n-3}.
\end{align}
2+1 dimensional gravity has given us many interesting results, we cannot just ignore that case. The static solution for $n=3$ is
\begin{align}
ds^{2}=- \Big (1-r_{0}-\frac{r^{2}}{l^{2}} \Big )dt^{2}+\Big (1-r_{0}-\frac{r^{2}}{l^{2}}\Big)^{-1}dr^{2}+r^{2}d\phi^{2}\label{3dmetric}
\end{align}
 for which one obtains from (\ref{yawru})  $E=r_{0}/2G_3$ again with $r_0$ a dimensionless constant and $G_3$ has inverse mass dimensions. 

\subsubsection{Energy and angular momentum of Kerr-AdS black holes in four dimensions }

Now let us use (\ref{newcharge2}) to calculate the conserved charges of the   Kerr-$AdS$ black hole in four dimensions. The original computation was done in \cite{emel1} which is pretty straightforward. Any form of the solution will do the job, but let us start with the Kerr-Schild form 

\begin{equation}
ds^{2}=d\bar{s}^{2}+\frac{2G M r }{\rho^2}k_{\mu}k_\nu dx^{\mu} dx^\nu, \hskip 1 cm \rho^{2} :=r^{2}+a^{2}\cos^{2}\theta,
\end{equation}
and the background global AdS metric is given as 
\begin{eqnarray}
d\bar{s}^{2} =  -\frac{\left(1-\frac{\Lambda r^{2}}{3}\right)\Delta_{\theta}dt^{2}}{\left(1+\frac{\Lambda a^{2}}{3}\right)}+\frac{\rho^{2}dr^{2}}{\left(1-\frac{\Lambda r^{2}}{3}\right)\left(r^{2}+a^{2}\right)}
 + \frac{\rho^{2}d\theta^{2}}{\Delta_{\theta}}+\frac{\left(r^{2}+a^{2}\right)\sin^{2}\theta d\phi^{2}}{\left(1+\frac{\Lambda a^{2}}{3}\right)},
\end{eqnarray}
where $\Delta_{\theta} = 1+ \frac{\Lambda}{3} a^2\cos^2 \theta$. 
The vector $k_{\mu}$ is null with respect to the full and background metrics and reads as 
\begin{equation}
k_{\mu}dx^{\mu}=\frac{\Delta_{\theta}dt}{\left(1+\frac{\Lambda a^{2}}{3}\right)}+\frac{\rho^{2}dr}{\left(1-\frac{\Lambda r^{2}}{3}\right)\left(r^{2}+a^{2}\right)}-\frac{a \sin^{2}\theta d\phi}{\left(1+\frac{\Lambda a^{2}}{3}\right)}.
\nonumber 
\end{equation}
The time-like  Killing vector is $ \bar \xi = ( -1,0,0,0)$, for which (\ref{newcharge2}) reduces to 
\begin{equation}
E = \frac{3}{ 16 \pi G \Lambda} \int_{S^2_\infty} d\Omega  (R^{r t}\thinspace_{\beta\sigma})^{\left(1\right)}\bar{\nabla}^\beta\bar{\xi}^\sigma,
\end{equation}
where the measure, including the normalizations of $\hat n_\mu$ and $\hat \sigma_\mu$ in $\bar\epsilon_{\mu\nu}$, contributes the factor $ \frac{r^2+ a^2 \cos^2\theta}{1+ \frac{\Lambda}{3} a^2}$ (with $d\Omega = \sin\theta\, d\theta\, d\phi$). The integral is over a sphere at $r \rightarrow \infty$ which yields the answer
\begin{equation}
E=\frac{M}{\left(1+\frac{\Lambda a^{2}}{3}\right)^{2}}.
\end{equation}
So the integration constant $M$ that appears during the integration of the field equations does not represent the total mass/energy of the Kerr-AdS black hole. The cosmological constant and the rotation parameter also enter the picture. 

To find the angular momentum, one uses the corresponding Killing vector $\bar \xi = ( 0,0,0,1)$, and (\ref{newcharge2}) yields 
\begin{equation}
J=\frac{aM}{\left(1+\frac{\Lambda a^{2}}{3}\right)^{2}}.
\end{equation}
So not only $a$ and $M$ conspire to form the angular momentum, but the cosmological constant also enters the picture. In the $\Lambda =0$ case, one arrives at the expected result. But even for generic $\Lambda$, one has the relation 
 $a E=J$.

$n$ dimensional version of this computation is a little bit cumbersome and it was done in \cite{8} in sufficient detail. See also \cite{emelson}.

\section{Charges of Quadratic Curvature Gravity}
This part will expound upon \cite{3,4}.  Quadratic gravity theory serves as a template to all other $f(R^{\mu\nu}_{\sigma\rho})$ type theories. So 
let us consider the generic quadratic gravity theory in $n$ dimensions with the action 
\begin{equation}
I=\int d^{n}x\sqrt{-g}\left(\frac{1}{\kappa}\left(R-2\Lambda_{0}\right)+\alpha R^{2}+\beta R_{\mu\nu}R^{\mu\nu}+\gamma\left(R_{\mu\nu\rho\sigma}R^{\mu\nu\rho\sigma}-4R_{\mu\nu}R^{\mu\nu}+R^{2}\right)\right),\label{eq:Quad_act}
\end{equation}
and the field equations
\begin{align}
&\frac{1}{\kappa}\left(R_{\mu\nu}-\frac{1}{2}g_{\mu\nu}R+g_{\mu\nu}\Lambda_{0}\right)+2\alpha R\left(R_{\mu\nu}-\frac{1}{4}g_{\mu\nu}R\right)+\left(2\alpha+\beta\right)\left(g_{\mu\nu}\Box-\nabla_{\mu}\nabla_{\nu}\right)R\nonumber \\
&+2\gamma\left(RR_{\mu\nu}-2R_{\mu\sigma\nu\rho}R^{\sigma\rho}+R_{\mu\sigma\rho\tau}R_{\nu}^{\phantom{\nu}\sigma\rho\tau}-2R_{\mu\sigma}R_{\phantom{\sigma}\nu}^{\sigma}-\frac{1}{4}g_{\mu\nu}\left(R_{\alpha\beta\rho\sigma}R^{\alpha\beta\rho\sigma}-4R_{\alpha\beta}R^{\alpha\beta}+R^{2}\right)\right)\nonumber \\
&+\beta\Box\left(R_{\mu\nu}-\frac{1}{2}g_{\mu\nu}R\right)+2\beta\left(R_{\mu\sigma\nu\rho}-\frac{1}{4}g_{\mu\nu}R_{\sigma\rho}\right)R^{\sigma\rho}  =\tau_{\mu\nu}.\label{eq:EoM_quad}
\end{align}
Our first task is to find the maximally symmetric vacua with $\tau_{\mu \nu}=0$. For this theory,  there are two maximally symmetric vacua of which the effective
cosmological constant $\Lambda$ are solutions to
\begin{equation}
\frac{\Lambda-\Lambda_{0}}{2\kappa}+k\Lambda^{2}=0,\qquad k:=\left(n\alpha+\beta\right)\frac{\left(n-4\right)}{\left(n-2\right)^{2}}+\gamma\frac{\left(n-3\right)\left(n-4\right)}{\left(n-1\right)\left(n-2\right)}.\label{eq:Eff_Lambda}
\end{equation}
It is already clear that $n=4$ stands out: it has a unique maximally symmetric vacuum for given $\Lambda_0$. Let us work in generic $n$ dimensions.
One of these two vacua must be chosen as the background spacetime about which we shall linearize. The procedure is the same as in Einstein's theory. The linearized field equations yield a fourth-order partial differential equation

\begin{equation}
{\mathcal{C}}\,\mathcal{G}_{\mu\nu}^{\left(1\right)}+\left(2\alpha+\beta\right)\left(\bar{g}_{\mu\nu}\bar{\square}-\bar{\nabla}_{\mu}\bar{\nabla}_{\nu}+\frac{2\Lambda}{n-2}\bar{g}_{\mu\nu}\right)R_{\left(1\right)}+\beta\left(\bar{\square}\mathcal{G}_{\mu\nu}^{\left(1\right)}-\frac{2\Lambda}{n-1}\bar{g}_{\mu\nu}R_{\left(1\right)}\right)=:T_{\mu\nu},\label{Linearized_eom}
\end{equation}
where $T_{\mu\nu}$ includes all the non-linear terms plus the bounded source $\tau_{\mu \nu}$. The constant ${\mathcal{C}}$ in-front of the linearized Einstein tensor
reads 
\begin{equation}
{\mathcal{C}}:=\frac{1}{\kappa}+\frac{4\Lambda n}{n-2}\alpha+\frac{4\Lambda}{n-1}\beta+\frac{4\Lambda\left(n-3\right)\left(n-4\right)}{\left(n-1\right)\left(n-2\right)}\gamma.\label{eq:c}
\end{equation}
Background covariant conservation of $T_{\mu\nu}$ is clear from our earlier generic construction, but we can also check this explicitly with  the following two identities together with  $\bar{\nabla}_{\mu}{\cal G}^{\left(1\right)\mu\nu}=0$:
\begin{align}
\bar{\nabla}^{\mu}\left(\bar{g}_{\mu\nu}\bar{\Box}-\bar{\nabla}_{\mu}\bar{\nabla}_{\nu}+\frac{2\Lambda}{n-2}\bar{g}_{\mu\nu}\right)R_{\left(1\right)} & =0,\nonumber \\
\bar{\nabla}^{\mu}\left(\bar{\Box}{\cal G}_{\mu\nu}^{\left(1\right)}-\frac{2\Lambda}{n-1}\bar{g}_{\mu\nu}R_{\left(1\right)}\right) & =0,
\end{align}
where $\bar{g}^{\mu\nu}{\cal G}_{\mu\nu}^{\left(1\right)}=\frac{2-n}{2}R_{\left(1\right)}$.
In writing  $\bar{\xi}_{\mu}T^{\mu\nu}$ as the divergence of an  anti-symmetric two tensor, one needs the following identities which are easy to prove
\begin{align}
&\bar{\xi}_{\nu}\bar{\Box}{\cal G}^{\left(1\right)\mu\nu}=  \bar{\nabla}_{\alpha}\Big(\bar{\xi}_{\nu}\bar{\nabla}^{\alpha}{\cal G}_{\left(1\right)}^{\mu\nu}-\bar{\xi}_{\nu}\bar{\nabla}^{\mu}{\cal G}_{\left(1\right)}^{\alpha\nu}-{\cal G}_{\left(1\right)}^{\mu\nu}\bar{\nabla}^{\alpha}\bar{\xi}_{\nu}+{\cal G}_{\left(1\right)}^{\alpha\nu}\bar{\nabla}^{\mu}\bar{\xi}_{\nu}\Big )\nonumber \\
 & \hskip 2.5 cm +{\cal G}_{\left(1\right)}^{\mu\nu}\Box\bar{\xi}_{\nu}+\bar{\xi}_{\nu}\bar{\nabla}_{\alpha}\bar{\nabla}^{\mu}{\cal G}_{\left(1\right)}^{\alpha\nu}-{\cal G}_{\left(1\right)}^{\alpha\nu}\bar{\nabla}_{\alpha}\bar{\nabla}^{\mu}\bar{\xi}_{\nu} ,\nonumber \\
&\bar{\nabla}_{\alpha}\bar{\nabla}_{\beta}\bar{\xi}_{\nu}=\bar{R}_{\phantom{\mu}\nu\beta\alpha}^{\mu}\bar{\xi}_{\mu}=\frac{2\Lambda}{\left(n-2\right)\left(n-1\right)}\left(\bar{g}_{\nu\alpha}\bar{\xi}_{\beta}-\bar{g}_{\alpha\beta}\bar{\xi}_{\nu}\right),\qquad\bar{\Box}\bar{\xi}_{\mu}=-\frac{2\Lambda}{n-2}\bar{\xi}_{\mu},\nonumber \\
&\bar{\xi}_{\nu}\bar{\nabla}_{\alpha}\bar{\nabla}^{\mu}{\cal G}_{\left(1\right)}^{\alpha\nu}=\frac{2\Lambda n}{\left(n-2\right)\left(n-1\right)}\bar{\xi}_{\nu}{\cal G}_{\left(1\right)}^{\mu\nu}+\frac{\Lambda}{n-1}\bar{\xi}^{\mu}R_{\left(1\right)}.
\end{align}
Therefore the conserved charges of quadratic
gravity for asymptotically (A)dS spacetimes read 
\begin{align}
Q_{{\rm QG}}\left(\bar{\xi}\right)= & \left( \mathcal{C}+\frac{4\Lambda\beta}{\left(n-1\right)\left(n-2\right)}\right)\int d^{n-1}x\sqrt{-\bar{g}}\bar{\xi}_{\nu}{\cal G}_{\left(1\right)}^{0\nu}\nonumber \\
 & +\left(2\alpha+\beta\right)\int_{\partial\bar{\Sigma}}dS_{i}\left(\bar{\xi}^{0}\bar{\nabla}^{i}R_{\left(1\right)}+R_{\left(1\right)}\bar{\nabla}^{0}\bar{\xi}^{i}-\bar{\xi}^{i}\bar{\nabla}^{0}R_{\left(1\right)}\right)\nonumber \\
 & +\beta\int_{\partial\bar{\Sigma}}dS_{i}\left(\bar{\xi}_{\nu}\bar{\nabla}^{i}{\cal G}_{\left(1\right)}^{0\nu}-\bar{\xi}_{\nu}\bar{\nabla}^{0}{\cal G}_{\left(1\right)}^{i\nu}-{\cal G}_{\left(1\right)}^{0\nu}\bar{\nabla}^{i}\bar{\xi}_{\nu}+{\cal G}_{\left(1\right)}^{i\nu}\bar{\nabla}^{0}\bar{\xi}_{\nu}\right),\label{fullcharge}
\end{align}
where the first line is, up to the factor $\kappa/\kappa_{\rm eff}$ defined below, the Einsteinian result (\ref{adt1}).  Let us carefully look at the second and third lines: 
for asymptotically
(A)dS spacetimes, the last two lines in (\ref{fullcharge}) vanish
since in these backgrounds ${\cal G}_{\mu\nu}^{\left(1\right)}\rightarrow O\left(r^{-n+1}\right)$, and $R_{(1)}$ falls off sufficiently fast as well. Therefore, the final result for generic quadratic gravity is a rather compact expression
\begin{tcolorbox}
\begin{equation}
\phantom{\frac{\frac{\xi}{\xi}}{\frac{\xi}{\xi}}}Q_{{\rm QG}}\left(\bar{\xi}\right)=\frac{\kappa}{\kappa_{{\rm eff}}}Q_{\text{Einstein}}\left(\bar{\xi}\right)
,\phantom{\frac{\frac{\xi}{\xi}}{\frac{\xi}{\xi}}}\label{eq:Q_quad}
\end{equation}
\end{tcolorbox}
where the effective Newton's constant reads
\begin{equation}
\frac{1}{\kappa_{\text{eff}}}:=\frac{1}{\kappa}+\frac{4\Lambda\left(n\alpha+\beta\right)}{n-2}+\frac{4\Lambda\left(n-3\right)\left(n-4\right)}{\left(n-1\right)\left(n-2\right)}\gamma={\mathcal{C}}+\frac{4\Lambda\beta}{\left(n-1\right)\left(n-2\right)}.\label{eq:kappa_eff_for_conserved_charges}
\end{equation}
So as far as asymptotically AdS spacetimes are considered, the effects of conserved charges are encoded in this effective Newton's constant.  But, for asymptotically flat backgrounds, $\kappa_{{\rm eff}}=\kappa$ and the higher order terms do not contribute to the charges, and the ADM results are intact. 

We can see the non-triviality of the above-conserved charges in the static solution of the $n$ dimensional static Boulware-Deser solution \cite{BDeser}  of the Einstein-Gauss-Bonnet theory.  For  $\alpha=\beta=\Lambda_{0}=0$ for the spherically symmetric ansatz
\begin{equation}
{\rm d}s^{2}=g_{00}{\rm d}t^{2}+g_{rr}{\rm d}r^{2}+r^{2}{\rm d}\Omega_{n-2},
\end{equation}
 the two solutions with different asymptotic structures  are given as 
\begin{equation}
-g_{00}=g_{rr}^{-1}=1+\frac{r^{2}}{2\kappa\gamma\left(n-3\right)\left(n-4\right)}\left(1\pm\left(1+4\kappa\gamma\left(n-3\right)\left(n-4\right)\frac{r_{0}^{n-3}}{r^{n-1}}\right)^{\frac{1}{2}}\right).\label{eq:BD_BH}
\end{equation}
The minus branch is the asymptotically flat Schwarzschild black hole; and
the plus branch is the asymptotically Schwarzschild-AdS
black hole. For large $r$, one has two different, and rather curious, behaviors:
\begin{equation}
-g_{00}\rightarrow1-\left(\frac{r_{0}}{r}\right)^{n-3},\qquad\qquad-g_{00}\rightarrow1+\left(\frac{r_{0}}{r}\right)^{n-3}+\frac{r^{2}}{\kappa\gamma\left(n-3\right)\left(n-4\right)},\label{eq:asymptotic}
\end{equation}
where the effective cosmological constant for the second branch is 
\begin{equation}
\Lambda=-\frac{\left(n-1\right)\left(n-2\right)}{2\kappa\gamma\left(n-3\right)\left(n-4\right)}.
\end{equation}
Observe that there is a minus sign in front of the second term in the asymptotically flat case, while the second term comes with a plus sign in the second branch. This ostensibly looks problematic as one of these branches might have a wrong sign for the energy and hence instability. This is not the case: the first branch, as already computed before, has the correct sign within the class of asymptotically flat spacetimes, while for the second branch, we have 
from (\ref{eq:Q_quad}) with $\kappa_{{\rm eff}}=-\kappa$; therefore,
\begin{equation}
E=\frac{\left(n-2\right)}{4G_{n}}r_{0}^{n-3},
\end{equation}
and the energy of the Boulware-Deser black hole is positive for both branches.

\section{Further remarks and extensions to generic gravity theories}

Let us make some remarks about the quadratic theory and its extensions. The details of the computations can be found in the relevant papers noted below.
\subsection{ Particle content of the quadratic theory}
 The quadratic theory (\ref{eq:Quad_act}) has 
 \begin{equation}
 n(n-2) = \frac{n(n-3)}{2} +\frac{(n+1)(n-2)}{2} +1
 \end{equation}
 degrees of freedom. The first number refers to the usual massless spin-2 graviton, the second one to the massive spin-2 graviton (generically a ghost), and the third one to a massive spin-0 scalar mode.  Note that counting the number of degrees of freedom from the metric and the gauge degrees of freedom alone is simply wrong. For example, in 2+1 dimensions Einstein's theory with or without a cosmological constant has zero bulk degrees of freedom, while the 2+1 dimensional quadratic gravity has generically 3 degrees of freedom. 
 
 Expansion of the quadratic action  (\ref{eq:Quad_act}) up to second order yields 
\begin{equation}
{\cal L}_2 = - \frac{1}{2} \Big ( {\mathcal{C}} + \frac{ 4 \Lambda \beta}{(n-1)(n-2)} \Big ) h^{\mu \nu} \mathcal{G}_{\mu\nu}^{(1)}  + \beta \mathcal{G}_{\mu\nu}^{(1)} \mathcal{G}^{\mu\nu}_{(1)} +
\Big ( \alpha + \frac{ \beta ( 4-n)}{4} \Big) R_{(1)}^2,
\label{orderh2}
\end{equation}
where the Riemann-square term nicely turns into gauge invariant combinations. All three mentioned modes are coupled here and diagonalization of these requires some work which was done in \cite{tekin_rapid}. From that computation follow the masses of the massive spin-2 and massive spin-0 modes, respectively, as
\begin{tcolorbox}
\begin{equation}
m_g^2 = -\frac{1}{\beta  \kappa }-4 \Lambda\frac{(n-1) (\beta +\alpha  n)+\gamma
   (n-4) (n-3)}{\beta  (n-2) (n-1)},
\end{equation}
\begin{equation}
m_s^2 = \frac{n-2}{\kappa  (4 \alpha  (n-1)+\beta
   n)}+ \frac{4 \Lambda (n-4) \Big((n-1) (\beta +\alpha  n)+\gamma  (n-3) (n-2) \Big)}{(n-1)
   (n-2) (4 \alpha  (n-1)+\beta  n)},
\end{equation}
\end{tcolorbox}
\noindent where $\Lambda$ is given as one of the solutions of (\ref{eq:Eff_Lambda}). In addition to these, there are the massless spin-2 modes. 
Let us make several remarks:
\begin{itemize}
\item One should not be surprised that all the parameters in the theory conspire to give the masses of the fundamental degrees of freedom.  For $\Lambda =0$, the computation of the Newtonian limit shows explicitly that there are two Yukawa-type interactions, one being repulsive, in addition to the usual Newtonian attraction and scalar attraction.  Generic propagator structure and tree-level particle exchange were computed in \cite{ahah}.
\item  Consider again the $\Lambda =0$ case, in generic $n$ dimensions, the massless spin-2 mode requires $\kappa >0$ to be a non-ghost ( i.e. with positive kinetic energy) and the massive spin-2 mode requires the opposite  $\kappa <0$ to be a non-ghost. There is no reconciliation of this except in $n=2+1$ dimensions for which the massless spin-2 mode does not exist, and the massive spin-0 mode decouples if $8 \alpha + 3\beta =0$. This is the New Massive Gravity \cite{nmg1} which has only a massive graviton with 2 degrees of freedom \cite{ahah,canon}. 

\item For $\alpha =\beta =0$, both of the massive modes decouple, and one has only a massless spin-2 mode in this Einstein-Gauss-Bonnet theory. 

\item The particle spectrum of a generic gravity theory with the Lagrangian $\mathcal{L}=\sqrt{-g}\,f\left(R_{\rho\sigma}^{\mu\nu}\right)$ can be found from the quadratic theory with the techniques discussed in the next subsection. See \cite{tekin_rapid} for details. 
\end{itemize}

\subsection{Energy and Angular Momentum in Generic  $\mathcal{L}=\sqrt{-g}\,f\left(R_{\rho\sigma}^{\mu\nu}\right)$
Theories}

Finding the conserved charges and the particle content of a generic gravity theory with a Lagrangian density of the form in the title of this section is a little non-trivial. There could be several ways to proceed, but let us adopt the technique that we have advocated in several works before, see for example  \cite{Sismanall} and \cite{Senturk} for the original expositions. Here we shall just give a summary of the final result as a recipe. Starting from the action

\begin{equation}
S=\int d^{n}x\,\sqrt{-g}\, f\left(R_{\rho\sigma}^{\mu\nu}\right),
\label{content1}
\end{equation}
one can find the matter-coupled field equations as
\begin{align}
\hspace*{-0.7cm}\frac{1}{2}\left(g_{\nu\rho}\nabla^{\lambda}\nabla_{\sigma}-g_{\nu\sigma}\nabla^{\lambda}\nabla_{\rho}\right)\frac{\partial f}{\partial R_{\rho\sigma}^{\mu\lambda}}-\frac{1}{2}\left(g_{\mu\rho}\nabla^{\lambda}\nabla_{\sigma}-g_{\mu\sigma}\nabla^{\lambda}\nabla_{\rho}\right)\frac{\partial f}{\partial R_{\rho\sigma}^{\lambda\nu}}\nonumber \\
-\frac{1}{2}\left(\frac{\partial f}{\partial R_{\rho\sigma}^{\mu\lambda}}R_{\rho\sigma\phantom{\lambda}\nu}^{\phantom{\rho\sigma}\lambda}-\frac{\partial f}{\partial R_{\rho\sigma}^{\lambda\nu}}R_{\rho\sigma\phantom{\lambda}\mu}^{\phantom{\rho\sigma}\lambda}\right)-\frac{1}{2}g_{\mu\nu}f & = \tau_{\mu \nu},\label{eq:Field_eqns_of_fRiem}
\end{align}
and linearize them about a generic background and follow the procedure outlined so far. Instead of that, let us discuss another way of carrying out the computation.  One can construct a {\it quadratic action in curvature} that has the same vacua, particle content, and linearized field equations as the original action (\ref{content1}). We call that beneficial action ``the equivalent quadratic action'' which reads as the Taylor series expansion of the original action   (\ref{content1}) around a maximally symmetric background up to second order in the curvature tensor:
\begin{align}
\hspace*{-0.5cm}S_{\text{EQA}}=\int d^{n}x\,\sqrt{-g} & \left\{ f\left(\bar{R}_{\alpha\beta}^{\mu\nu}\right)+\left[\frac{\partial f}{\partial R_{\rho\sigma}^{\lambda\nu}}\right]_{\bar{R}_{\rho\sigma}^{\mu\nu}}\left(R_{\rho\sigma}^{\lambda\nu}-\bar{R}_{\rho\sigma}^{\lambda\nu}\right)\right.\nonumber \\
 & \left.+\frac{1}{2}\left[\frac{\partial^{2}f}{\partial R_{\alpha\tau}^{\eta\theta}\partial R_{\rho\sigma}^{\mu\lambda}}\right]_{\bar{R}_{\rho\sigma}^{\mu\nu}}\left(R_{\alpha\tau}^{\eta\theta}-\bar{R}_{\alpha\tau}^{\eta\theta}\right)\left(R_{\rho\sigma}^{\mu\lambda}-\bar{R}_{\rho\sigma}^{\mu\lambda}\right)\right\} .\label{eq:EQCA_as_Taylor_epansion}
\end{align}
So our full action $S$ and the beneficial action $S_{\text{EQA}}$  are the same theories up to the second order in the curvature. One can use this fact to find the particle content and the conserved charges of the full theory. For this purpose, given the action, namely, $f\left(R_{\rho\sigma}^{\mu\nu}\right)$, one should calculate the following quantities

\begin{align}
\left[\frac{\partial f}{\partial R_{\rho\sigma}^{\mu\nu}}\right]_{\bar{R}_{\rho\sigma}^{\mu\nu}}R_{\rho\sigma}^{\mu\nu}  &=:\zeta R,\label{eq:First_order}\\
\frac{1}{2}\left[\frac{\partial^{2}f}{\partial R_{\rho\sigma}^{\mu\nu}\partial R_{\lambda\gamma}^{\alpha\beta}}\right]_{\bar{R}_{\rho\sigma}^{\mu\nu}}R_{\rho\sigma}^{\mu\nu}R_{\lambda\gamma}^{\alpha\beta} &=:\alpha R^{2}+\beta R_{\sigma}^{\lambda}R_{\lambda}^{\sigma}+\gamma\left(R_{\rho\sigma}^{\mu\nu}R_{\mu\nu}^{\rho\sigma}-
4R_{\nu}^{\mu}R_{\mu}^{\nu}+R^{2}\right).\label{eq:Second_order}
\end{align}
So the parameters $\zeta$, $\alpha$, $\beta$, $\gamma$ are to be determined
from these equations.  $\alpha$, $\beta$, and $\gamma$ will appear exactly in the equivalent quadratic action (\ref{eq:EQCA_as_Taylor_epansion}), and the remaining two parameters come from the following two equations 
\begin{align}
\frac{1}{\kappa} & =\zeta-\left(\frac{4\Lambda}{n-2}\left(n\alpha+\beta\right)+\frac{4\Lambda\left(n-3\right)}{n-1}\gamma\right), \nonumber\\
\frac{\Lambda_{0}}{\kappa} & =-\frac{1}{2}f\left(\bar{R}_{\rho\sigma}^{\mu\nu}\right)+\frac{\Lambda n}{n-2}\zeta-\frac{2\Lambda^{2}n}{\left(n-2\right)^{2}}\left(n\alpha+\beta\right)-\frac{2\Lambda^{2}n\left(n-3\right)}{\left(n-1\right)\left(n-2\right)}\gamma.\label{Lameff}
\end{align}
Then, the gravitational charges of the $f\left(R_{\rho\sigma}^{\mu\nu}\right)$
the theory is given as
\begin{align}
Q_{f\left(R_{\rho\sigma}^{\mu\nu}\right)}(\bar{\xi})= & \left(\frac{1}{\kappa}+\frac{4\Lambda n}{n-2}\alpha+\frac{4\Lambda}{n-2}\beta+\frac{4\Lambda\left(n-3\right)\left(n-4\right)}{\left(n-1\right)\left(n-2\right)}\gamma\right)Q_{\text{Einstein}}(\bar{\xi}).\label{eq:Q_F}
\end{align}
Here  $Q_{\text{Einstein}}(\bar{\xi})$ is computed with $\kappa =1$. Here, note that  $\alpha$, $\beta$, $\gamma$, $\kappa$ are to be found from (\ref{eq:First_order}-\ref{Lameff}), and the effective
cosmological constant $\Lambda$ satisfies (\ref{quadratic}).  See \cite{Amsel} for a similar formulation of conserved charges in generic $f\left(R_{\rho\sigma}^{\mu\nu}\right)$ theories.

\subsubsection{ Born-Infeld extension of New Massive Gravity as an example}
As an example of the above construction, let us give the conserved charges of the BTZ \cite{BTZ} black hole in the  Born-Infeld New Massive Gravity (BINMG) \cite{BINMG1,BINMG2} defined by the action
\begin{equation}
I_{\text{BINMG}}=-4m^{2}\int d^{3}x\,\left[\sqrt{-\det\left(g_{\mu\nu}+\frac{\sigma}{m^{2}}G_{\mu\nu}\right)}-\left(1-\frac{\lambda_{0}}{2}\right)\sqrt{-g}\right],\label{eq:det_BINMG_action}
\end{equation}
where $G_{\mu \nu}:= R_{\mu \nu} - \frac{1}{2}R g_{\mu \nu}$ and we defined a dimensionless bare cosmological constant $\lambda_0 = \Lambda_0/m^2$; we have a choice of sign $\sigma = \pm1$. This theory has several interesting properties:
\begin{enumerate}
\item For $\lambda_{0}\ne0$, unlike any generic finite order theory besides the cosmological Einstein's theory, it has a unique maximally symmetric vacuum with  an effective cosmological constant $\Lambda = \lambda m^2$ given as  \cite{Nam,BINMG2}
\begin{equation}
\lambda=\sigma \lambda_{0}\left(1-\frac{\lambda_{0}}{4}\right),\quad\lambda_{0}<2.
\end{equation}
Flat space is the unique vacuum when $\lambda_{0}=0$. The uniqueness of the vacuum is extremely important as we have seen above: there is no classical argument to pick one vacuum over another. 
\item For $\sigma=-1$, it has a unitary spin-2 massive degree of freedom with $m_g^{2}=m^{2}(1+\lambda)$
about the flat ($\Lambda=0$) and AdS backgrounds. [For generic $\sigma$, the mass formula of the previous subsection gives $m_g^{2}=m^{2}(\lambda-\sigma)$.] This provides an infinite-order extension of the quadratic NMG
which has the same perturbative properties.

\item It reproduces, up to desired order in the curvature expansion, the extended NMG theories that are consistent with the AdS/CFT duality and that have a viable $c$-function \cite{Sinha,BINMG2,Paulos}.
\item The BINMG action appears as a counterterm in $\text{AdS}{}_{4}$ \cite{Jatkar}.
\item It might have a supersymmetric extension \cite{Ozkan}.
\end{enumerate}

To apply the method laid out above, the quantities we need to calculate are :
\begin{align}
&f\left(\bar{R}_{\nu}^{\mu}\right)  =4m^{2}\left[\left(1-\frac{\lambda_{0}}{2}\right)-\left(1-\sigma\lambda\right)^{3/2}\right],\nonumber \\
&\left[\frac{\partial f}{\partial R_{\beta}^{\alpha}}\right]_{\bar{R}_{\nu}^{\mu}}R_{\beta}^{\alpha}  =\sigma\left(1-\sigma\lambda\right)^{1/2}R,\label{eq:}\\
&\frac{1}{2}\left[\frac{\partial^{2}f}{\partial R_{\sigma}^{\rho}\partial R_{\beta}^{\alpha}}\right]_{\bar{R}_{\nu}^{\mu}}R_{\sigma}^{\rho}R_{\beta}^{\alpha}  =\frac{1}{m^{2}}\left(1-\sigma\lambda\right)^{-1/2}\left(R_{\nu}^{\mu}R_{\mu}^{\nu}-\frac{3}{8}R^{2}\right),\nonumber 
\end{align}
where $\sigma\lambda<1$ should
be satisfied. From these, one can simply read the effective parameters
of the equivalent quadratic action as
\begin{equation}
\zeta=\sigma\left(1-\sigma\lambda\right)^{1/2},\qquad\beta=-\frac{8}{3}\alpha=\frac{1}{m^{2}}\left(1-\sigma\lambda\right)^{-1/2}.
\end{equation}
Using this result in  (\ref{Lameff}),
one gets 
\begin{align}
\frac{1}{\kappa} =\frac{\left(\sigma-\frac{\lambda}{2}\right)}{\sqrt{1-\sigma\lambda}},\hskip 1 cm 
\frac{\Lambda_{0}}{\kappa} =m^{2}\left[\lambda_{0}-2+\frac{1}{\sqrt{1-\sigma\lambda}}\left(2-\sigma\lambda-\frac{\lambda^{2}}{4}\right)\right].\label{eq:BINMG_kappa_Lambda_0}
\end{align}
 In \cite{Nam,BINMG2}, it was shown that the BTZ black
hole 
\begin{equation}
ds^{2}=-N^{2}dt^{2}+N^{-2}dr^{2}+r^{2}\left(N^{\phi}dt+d\phi\right)^{2},\label{eq:BTZ_metric}
\end{equation}
with the metric functions
\begin{equation}
N^{2}\left(r\right)=-M+\frac{r^{2}}{\ell^{2}}+\frac{J^{2}}{4r^{2}},\qquad N^{\phi}\left(r\right)=-\frac{J}{2r^{2}},
\end{equation}
is a solution to BINMG theory under the condition 
\begin{equation}
\lambda=\sigma\lambda_{0}\left(1-\frac{\lambda_{0}}{4}\right),\quad\lambda_{0}<2.\label{eq:BTZ_BINMG}
\end{equation}
Using (\ref{eq:Q_F}) the mass and the angular
momentum of the rotating  BTZ black hole in BINMG can be found as
\begin{equation}
E=\sigma\sqrt{1-\sigma\lambda}M,\qquad L=\sigma\sqrt{1-\sigma\lambda}J.\label{eq:E-L}
\end{equation}
Observe that as expected from (\ref{eq:Q_F}) the charges are scaled
yet their ratio is intact.

\section{Conclusions and Discussions}
We have given a somewhat detailed account of the Killing charge construction in generic gravity theories. We have not intended to present a novel or complete work here, but just to expand on our earlier work. In this section let us note some extensions and missing parts to the work presented here.
\begin{itemize}
\item For other approaches to conserved charges see \cite{Magnon,HawkingHorowitz,97,Aros,40,Hollands}
 and two beautiful expositions \cite{compere,Banados_rev}.
\item For conserved charges of quadratic theory in generic backgrounds see \cite{Devecioglu}.
\item For conserved charges in non-minimally coupled scalar tensor theory see \cite{9}.

\item For conserved charges in asymptotically AdS backgrounds of the topologically massive gravity \cite{djt} theory in 2+1 dimensions, see \cite{6} and its extension to generic backgrounds see \cite{clement}. Construction of these charges for generic backgrounds using the symplectic structure of the theory was carried out in \cite{Nazaroglu}. This construction follows the so-called ``covariant canonical formalism'' \cite{crnkovic}.

\item In theories with fermions one has to work within the vierbein and spin-connection formulation, the corresponding conserved charges in this formulation were given in \cite{Cebeci}. In \cite{Baykal} a detailed exposition of linearized gravity in terms of differential forms is given. 

\item For the construction of non-stationary energy on an initial data surface see \cite{Dain} and \cite{Kroon,non,Kroon2}. This construction could be relevant to gravitational wave physics, as it gives a refinement of the ADM energy and identifies the non-stationary part of the total energy using approximate Killing initial data.

\item In non-linear theories, such as all the gravity theories discussed here, the validity of the perturbation theory is not guaranteed. There are typically constraints on the first-order perturbation theory coming from the second-order perturbation theory. This is a subtle subject relevant to the so-called Taub \cite{Taubb} charges studied in \cite{Taub} and the references there. See also a detailed account of linearization instability in the Ph.D. thesis \cite{altas_tez}.

\item We studied diffeomorphisms here, but for large gauge transformations that modify the values of the charges, one should look at \cite{review} and the references therein.  

\item A similar construction of conserved charges carried out here can be worked out for Yang-Mills theories both in flat  \cite{yang-ab} and curved backgrounds \cite{ercan}. In the Yang-Mills case, the symmetries of the background field must be carefully defined.

\item  Let us tie our apparently {\it perturbative} energy treatment to the exact, fully non-linear treatment. 
Regge and Teitelboim \cite{Regge}, following a remark by DeWitt \cite{DeWitt}, showed that in the fully nonlinear Hamiltonian treatment of GR in spatially open manifolds, one has to include a boundary term $E[g]$ to the bulk Hamiltonian.  This comes from the fact that the functional derivatives of the functionals with respect to the canonical fields and momenta do not make sense otherwise, since one has to make integration by parts and pick up boundary terms which are not allowed in the definition of the functional derivative. Therefore, to derive Einstein's second-order equations from Hamilton's first-order equations,  one must add a surface term to the Hamiltonian which, on appearance, does not modify  Hamilton's equations but makes them legitimate,i.e. mathematically well-defined. That added boundary term is exactly the ADM energy ($E[g] = E_{\text{ADM}}[h]$). Moreover, the {\it value} of the full Hamiltonian, say $H$, which is a sum of the bulk and boundary terms, yields $H  = E_{ADM}[h]$  upon the use of the field equations. Namely, the linear-looking ADM energy captures all the nonlinear energy stored in the gravitational field and the localized matter in the bulk of the spacetime as long as field equations are satisfied.
\end{itemize}
\newpage

\section{Appendix}

\subsection{Curvature tensors at second order in perturbation theory}

Let us collect some formulas related to the perturbation theory applied in the bulk of the work.  Here we introduce a small parameter $\epsilon$ (instead of the $\kappa$ we used in the text) that counts the order then as an exact expression let us define
the metric perturbation $h_{\mu\nu}$
as \begin{equation}
g_{\mu\nu}=:\bar{g}_{\mu\nu}+\epsilon h_{\mu\nu}\label{eq:Perturbation_def}
\end{equation}
where $\bar{g}_{\mu\nu}$  is a generic background that need not be maximally symmetric for the ensuing discussion. Any quantity with a bar refers to the background metric. Raising and lowering the indices at the end are done with the background metric. 
The inverse metric $g^{\mu\nu}$ follows  as\begin{equation}
g^{\mu\nu}=\bar{g}^{\mu\nu}-\epsilon h^{\mu\nu}+\epsilon^{2}h^{\mu\rho}h_{\rho}^{\nu}+O\left(\epsilon^{3}\right).\label{eq:Oh2_of_inv_met}\end{equation}
 The trace of the metric perturbation is defined as $h:= \bar{g}^{\mu\nu}h_{\mu\nu}$.
In this order, the Christoffel reads as \begin{equation}
\Gamma_{\mu\nu}^{\rho}=\bar{\Gamma}_{\mu\nu}^{\rho}+\epsilon \left(\Gamma_{\mu\nu}^{\rho}\right)_{L}-\epsilon^{2}h_{\beta}^{\rho}\left(\Gamma_{\mu\nu}^{\beta}\right)_{L}+O\left(\epsilon^{3}\right),\end{equation}
 where $\bar{\Gamma}_{\mu\nu}^{\rho}$ is the background metric compatible
connection $\bar{\nabla}_{\rho}\bar{g}_{\mu\nu}=0$ and the linearized
connection $\left(\Gamma_{\mu\nu}^{\rho}\right)_{L}$ is defined to be the first order term as \begin{equation}
\left(\Gamma_{\mu\nu}^{\rho}\right)_{L}\:=\frac{1}{2}\bar{g}^{\rho\lambda}\Big (\bar{\nabla}_{\mu}h_{\nu\lambda}+\bar{\nabla}_{\nu}h_{\mu\lambda}-\bar{\nabla}_{\lambda}h_{\mu\nu}\Big ).\label{eq:Linear_Christoffel}\end{equation}
Here the letter $L$, be it subscript or superscript, refers to the linearized forms of the corresponding tensors. Now we have to find the Riemann tensor in the second order, for this purpose let us use 
\begin{equation}
\Gamma_{\mu\nu}^{\rho}=\bar{\Gamma}_{\mu\nu}^{\rho}+\delta\Gamma_{\mu\nu}^{\rho},
\end{equation}
in the Riemann tensor
\begin{equation}
R_{\phantom{\mu}\nu\rho\sigma}^{\mu}=\partial_{\rho}\Gamma_{\sigma\nu}^{\mu}+\Gamma_{\rho\lambda}^{\mu}\Gamma_{\sigma\nu}^{\lambda}-\rho\leftrightarrow\sigma
\end{equation}
to get
 \begin{equation}
R_{\phantom{\mu}\nu\rho\sigma}^{\mu}=\bar{R}_{\phantom{\mu}\nu\rho\sigma}^{\mu}+\bar{\nabla}_{\rho}\left(\delta\Gamma_{\sigma\nu}^{\mu}\right)-\bar{\nabla}_{\sigma}\left(\delta\Gamma_{\rho\nu}^{\mu}\right)+\delta\Gamma_{\rho\lambda}^{\mu}\delta\Gamma_{\sigma\nu}^{\lambda}-\delta\Gamma_{\sigma\lambda}^{\mu}\delta\Gamma_{\rho\nu}^{\lambda}.\end{equation}
The first and second-order terms are
\begin{equation}
\delta\Gamma_{\mu\nu}^{\rho}=\epsilon\left(\Gamma_{\mu\nu}^{\rho}\right)_{L}-\epsilon^{2}h_{\beta}^{\rho}\left(\Gamma_{\mu\nu}^{\beta}\right)_{L}.
\end{equation}
Therefore, the Riemann tensor up to the second order becomes a somewhat cumbersome expression
\begin{align}
R_{\phantom{\mu}\nu\rho\sigma}^{\mu}= & \bar{R}_{\phantom{\mu}\nu\rho\sigma}^{\mu}+\epsilon \left(R_{\phantom{\mu}\nu\rho\sigma}^{\mu}\right)_{L}-\epsilon^{2}h_{\beta}^{\mu}\left(R_{\phantom{\mu}\nu\rho\sigma}^{\beta}\right)_{L}\nonumber \\
 & -\epsilon^{2}\bar{g}^{\mu\alpha}\bar{g}_{\beta\gamma}\left[\left(\Gamma_{\rho\alpha}^{\gamma}\right)_{L}\left(\Gamma_{\sigma\nu}^{\beta}\right)_{L}-\left(\Gamma_{\sigma\alpha}^{\gamma}\right)_{L}\left(\Gamma_{\rho\nu}^{\beta}\right)_{L}\right]+O\left(\epsilon^{3}\right),\label{eq:Second_order_Riemann}\end{align}
with the linearized Riemann tensor given as 
\begin{align}
\left(R_{\phantom{\mu}\nu\rho\sigma}^{\mu}\right)_{L}=\frac{1}{2} & \Bigg ( \bar{\nabla}_{\rho}\bar{\nabla}_{\sigma}h_{\nu}^{\mu}+\bar{\nabla}_{\rho}\bar{\nabla}_{\nu}h_{\sigma}^{\mu}-\bar{\nabla}_{\rho}\bar{\nabla}^{\mu}h_{\sigma\nu}-\bar{\nabla}_{\sigma}\bar{\nabla}_{\rho}h_{\nu}^{\mu}
& -\bar{\nabla}_{\sigma}\bar{\nabla}_{\nu}h_{\rho}^{\mu}+\bar{\nabla}_{\sigma}\bar{\nabla}^{\mu}h_{\rho\nu}\Bigg).\label{eq:Linear_Riemann}\end{align}
Then  the Ricci tensor at second order is
 \begin{align}
R_{\nu\sigma}= & \bar{R}_{\nu\sigma}+\epsilon \left(R_{\nu\sigma}\right)_{L}-\epsilon^{2}h_{\beta}^{\mu}\left(R_{\phantom{\mu}\nu\mu\sigma}^{\beta}\right)_{L}\nonumber \\
 & -\epsilon^{2}\bar{g}^{\mu\alpha}\bar{g}_{\beta\gamma}\Bigg (\left(\Gamma_{\mu\alpha}^{\gamma}\right)_{L}\left(\Gamma_{\sigma\nu}^{\beta}\right)_{L}-\left(\Gamma_{\sigma\alpha}^{\gamma}\right)_{L}\left(\Gamma_{\mu\nu}^{\beta}\right)_{L}\Bigg )+O\left(\epsilon^{3}\right),\label{eq:Second_order_Ricci}\end{align}
and the scalar curvature in second order is
 \begin{align}
R= & \bar{R}+\epsilon R_{L}+\epsilon^{2}\left\{ \bar{R}^{\rho\lambda}h_{\alpha\rho}h_{\lambda}^{\alpha}-h^{\nu\sigma}\left(R_{\nu\sigma}\right)_{L}-\bar{g}^{\nu\sigma}h_{\beta}^{\mu}\left(R_{\phantom{\mu}\nu\mu\sigma}^{\beta}\right)_{L}\right.\nonumber \\
 & \phantom{\bar{R}+\tau^{2}R_{L}+\tau^{2}}\left.-\bar{g}^{\nu\sigma}\bar{g}^{\mu\alpha}\bar{g}_{\beta\gamma}\left[\left(\Gamma_{\mu\alpha}^{\gamma}\right)_{L}\left(\Gamma_{\sigma\nu}^{\beta}\right)_{L}-\left(\Gamma_{\sigma\alpha}^{\gamma}\right)_{L}\left(\Gamma_{\mu\nu}^{\beta}\right)_{L}\right]\right\},\label{eq:Second_order_R}\end{align}
where the linearized Ricci tensor and the linearized scalar curvature are defined, respectively, as \begin{equation}
R_{\nu\sigma}^{L}\equiv\frac{1}{2}\left(\bar{\nabla}_{\mu}\bar{\nabla}_{\sigma}h_{\nu}^{\mu}+\bar{\nabla}_{\mu}\bar{\nabla}_{\nu}h_{\sigma}^{\mu}-\bar{\Box}h_{\sigma\nu}-\bar{\nabla}_{\sigma}\bar{\nabla}_{\nu}h\right),\label{eq:Linear_Ricci}\end{equation}
 \begin{equation}
R_{L}=\bar{g}^{\alpha\beta}R_{\alpha\beta}^{L}-\bar{R}^{\alpha\beta}h_{\alpha\beta}.\label{eq:Linear_R}\end{equation}

\subsection{Diffeomorphism invariance in a geometric way }

This section follows \cite{Taubb} and appeared in \cite{Taub}.  We present a compact way of finding out the consequences of diffeomorphisms. 
Let $\lambda\in\mathbb{R}$ and $\varphi$ be a one-parameter family of diffeomorphisms acting on the spacetime manifold:
\begin{equation}
\varphi:\mathbb{R}\times\mathscr{M}\rightarrow\text{\ensuremath{\mathscr{M}}},
\end{equation}
then diffeomorphism invariance of a tensor field of generic rank $T$ means exactly the following statement
\begin{equation}
T(\varphi^{*}g)=\varphi^{*}T(g),\label{diffeomorphisminvariance}
\end{equation}
where $\varphi^{*}$ is the pullback map.  Let us denote the diffeomorphism by $\varphi_{\lambda}$ and assuming $\varphi_{0}$
to be the identity diffeomorphism/map, differentiating (\ref{diffeomorphisminvariance}) with respect to
$\lambda$ yields
\begin{equation}
\frac{d}{d\lambda}T(\varphi_\lambda^{*}g)=\frac{d}{d\lambda}\varphi_\lambda^{*}T(g).
\end{equation}
Using the chain rule one has 
\begin{equation}
DT(\varphi_{\lambda}^{*}g)\cdot\frac{d}{d\lambda}\varphi_{\lambda}^{*}g=\varphi_{\lambda}^{*}\left(\mathscr{L}_{X}T(g)\right),\label{firstderivative}
\end{equation}
where $D$ denotes the Fr\'echet derivative and $\mathscr{L}_{X}$ denotes the Lie derivative along the vector field $X$ that generates the diffeomorphism. In local coordinates for a rank $(0,2)$ tensor field-which is relevant for field equation-the last expression
yields
\begin{equation}
\delta_{X}(T_{\mu\nu})^{\left(1\right)}\cdot h=\mathscr{L}_{X}\bar{T}_{\mu\nu}.
\end{equation}
Specifically for the cosmological Einstein tensor $T_{\mu\nu}={\cal {G}}_{\mu\nu}= G_{\mu \nu} + \Lambda g_{\mu \nu}$, we have 
\begin{equation}
\delta_{X}({\cal {G}}_{\mu\nu})^{\left(1\right)}\cdot h=\mathscr{L}_{X}\bar{{\cal {G}}}_{\mu\nu}=0,
\end{equation}
which is a statement of the gauge invariance of the first-order linearized cosmological Einstein tensor.  

For the second-order tensors, we can take another derivative of (\ref{firstderivative}):
\begin{equation}
D^{2}T(g)\cdot\left(h,\text{\ensuremath{\mathscr{L}}}_{X}g\right)+DT(g)\cdot\text{\ensuremath{\mathscr{L}}}_{X}h=\text{\ensuremath{\mathscr{L}}}_{X}(DT(g)\cdot h),
\end{equation}
which yields in local coordinates 
\begin{equation}
\delta_{X}(T_{\mu\nu})^{\left(2\right)}\cdot[h,h]+(T_{\mu\nu})^{\left(1\right)}\cdot\mathscr{L}_{X}h=\text{\ensuremath{\mathscr{L}}}_{X}(T_{\mu\nu})^{\left(1\right)}\cdot h.
\end{equation}
 When $T_{\mu\nu}={\cal {G}}_{\mu\nu}$, we obtain
\begin{equation}
\delta_{X}({\cal {G}}_{\mu\nu})^{\left(2\right)}\cdot[h,h]+({\cal {G}}_{\mu\nu})^{\left(1\right)}\cdot\mathscr{L}_{X}h=\text{\ensuremath{\mathscr{L}}}_{X}({\cal {G}}_{\mu\nu})^{\left(1\right)}\cdot h.
\end{equation}
The right-hand side is zero for linearized solutions, and one obtains 
\begin{equation}
\delta_{X}({\cal {G}}_{\mu\nu})^{\left(2\right)}\cdot[h,h]=-({\cal {G}}_{\mu\nu})^{\left(1\right)}\cdot\mathscr{L}_{X}h.
\end{equation}
The right-hand side of this expression is not zero but it can be written as a pure divergence term.

\section{Acknowledgments and a disclaimer}
Over the years I have talked with many colleagues on issues discussed in this tribute. I cannot list them all here, but I thank them all. 
In particular for the preparation of this manuscript and the ideas in it; I would like to thank M. Gurses, A. Karasu, T.C. Sisman, I. Gullu, E. Altas, O. Sarioglu, H. Cebeci, E. Kilicarslan, H. Adami, M. Setare (whom we lost in a tragic accident in 2022), Y. Nutku (late), C. Nazaroglu, A.N. Petrov, S. M. Kopeikin, R. R. Lompay, M. Guler and A. Tavlayan. I would like to thank A. Waldron for reminding me of a reference that was particularly important for S. Deser, and for hosting me at Davis some time ago. 
 
 As a disclaimer let me note that the book \cite{book} and the review article \cite{review} include most of the material presented here: the only possible novelty is that I give more details of some of the computations.  I am also grateful to my former student T.C. Sisman for providing me with some of the details of our published work.

\end{document}